\documentclass[12pt]{article}
\usepackage{epsfig,amsfonts,amssymb}
\usepackage{hyperref}
\usepackage{cite}
\input epsf.sty
\usepackage{cite}

\def\ZZZ{{\hbox{ Z\kern-1.6mm Z}}}
\def\RRR{{\hbox{ R\kern-2.4mm R}}}
\def\CCC{{\hbox{ C\kern-2.0mm C}}}
\def\zzz{{\hbox{z\kern-1mm z}}}

\newcommand{\nn}{\nonumber \\}

\newcommand{\vt}{\vartheta}

\newcommand{\qeq}{{\hbox{=\kern-2.3mm ? \kern.5mm }}}
\renewcommand{\qeq}{=}

\newcommand{\eps}{\epsilon}

\newcommand{\ws}{{\wt\sigma}}
\newcommand{\wrh}{{\wt\rho}}
\newcommand{\wv}{{\wt v}}

\newcommand{\DD}{{\cal D}}

\newcommand{\CC}{{\cal C}}

\newcommand{\OO}{{\cal O}}

\newcommand{\LL}{{\cal L}}

\newcommand{\wt}{\widetilde}
\newcommand{\wh}{\widehat}

\newcommand{\NN}{{\cal N}}

\newcommand{\SSS}{{\cal S}}

\newcommand{\be}{\begin{equation}}
\newcommand{\ee}{\end{equation}}
\newcommand{\ben}{\begin{eqnarray}\displaystyle}
\newcommand{\een}{\end{eqnarray}}

\newcommand{\refb}[1]{(\ref{#1})}
\newcommand{\p}{\partial}
\newcommand{\sectiono}[1]{\section{#1}\setcounter{equation}{0}}

\def\one{{\hbox{ 1\kern-.8mm l}}}
\def\zero{{\hbox{ 0\kern-1.5mm 0}}}

\newcommand{\bea}[1]{\begin{eqnarray}\label{#1} }
\newcommand{\eea}{\end{eqnarray}}

\newcommand{\wJ}{\wt J}
\newcommand{\bN}{{\bf N}}

\begin{document}

\baselineskip 24pt

\begin{center}
{\Large \bf  Logarithmic Corrections to Rotating 
Extremal Black Hole
Entropy in Four and Five Dimensions}

\end{center}

\vskip .6cm
\medskip

\vspace*{4.0ex}

\baselineskip=18pt

\centerline{\large \rm Ashoke Sen}

\vspace*{4.0ex}

\centerline{\large \it Harish-Chandra Research Institute}
\centerline{\large \it  Chhatnag Road, Jhusi,
Allahabad 211019, India}

\vspace*{1.0ex}
\centerline{\small E-mail:  sen@mri.ernet.in}

\vspace*{5.0ex}

\centerline{\bf Abstract} \bigskip

We compute logarithmic
corrections to the entropy of rotating extremal black holes 
using quantum entropy function i.e. Euclidean quantum
gravity approach.
Our analysis includes five dimensional
supersymmetric BMPV black holes in 
type IIB string theory on $T^5$ and $K3\times S^1$ as well as in
the
five dimensional
CHL models, and also
non-supersymmetric extremal Kerr black hole and slowly
rotating extremal Kerr-Newmann black holes in four dimensions.
For BMPV black holes our results are in perfect
agreement with the microscopic results
derived from string theory. In particular
we reproduce correctly the dependence of
the logarithmic corrections on the number of U(1) gauge fields
in the theory, and on the angular momentum carried by the black
hole in different scaling limits. We also explain the 
shortcomings of the Cardy limit in explaining the
logarithmic corrections in the limit in which the (super)gravity description
of these black holes becomes a valid approximation.
For non-supersymmetric extremal black holes, {\it e.g.} for the
extremal Kerr black hole in four dimensions, our result provides a
stringent testing ground for
any microscopic explanation of the
black hole entropy, {\it e.g.} Kerr/CFT correspondence.

\vfill \eject

\baselineskip=18pt

\tableofcontents

\sectiono{Introduction and summary}  \label{sintro}

Supersymmetric extremal black holes enjoy a certain set of
non-renormalization properties, and this makes them
a very useful
testing ground for comparing the macroscopic predictions for the
entropy against the microscopic entropy\cite{9601029,9602065}. 
In particular
for a class of $\NN=4$ supersymmetric string
theories in four dimensions one can use Wald's 
formula\cite{9307038,9312023,9403028,9502009}
adapted to BPS black holes\cite{9812082,9904005,0007195}
for computing higher derivative corrections to the black hole
entropy, and the results are in  remarkable agreement
with the microscopic 
results\cite{9607026,0412287,0505094,0506249,
0508174,0510147,
0602254,0603066,0605210,0607155,0609109}.
Nevertheless, 
as there have been
other attempts to explain extremal 
black hole entropy without making
direct use of string theory\cite{9712251,0809.4266}, 
it is useful to explore to what
extent string theory can do better than these general
methods, particularly
in situations where quantum gravity corrections to black
hole entropy become
important. 

A generalization of Wald's formula,
based on path integral over various fields in
the near horizon geometry of the black hole, can be used to
compute quantum corrections to the extremal black hole 
entropy\cite{0809.3304,1008.3801}.\footnote{As has already been emphasized in
the past, 
extremal black holes refer to extremal limit of non-extremal black holes.
On the macroscopic side this is apparent from the form of the
Euclidean $AdS_2$ metric \refb{emets}, which is an analytic continuation
of the Lorentzian near horizon metric $- (r^2-1) dt^2 + dr^2/(r^2-1) $
($r=\cosh\eta$, $t=-i\theta$). This near horizon
geometry arises in the zero temperature
{\it limit} of black holes and still has a bifurcate Killing horizon at $r=1$.
The gravity / string theory partition function in this geometry is
then compared with the microscopic partition function
$Tr(e^{-H/T})$ in the zero temperature limit, which, for a gapped system,
approaches $d_0 e^{-E_0/T}$ where $E_0$ is the ground state energy
and $d_0$ is the ground state degeneracy.}
Recently this method has been used to compute logarithmic
corrections to the entropy in a class of four dimensional
$\NN\ge 2$ supersymmetric
string theories\cite{1005.3044,1106.0080,1108.3842,1109.0444}. 
These corrections are particularly
interesting since unlike higher derivative corrections which
are highly sensitive to the specific
string theory under consideration and
may not exist {\it e.g.} in a different approach to quantizing gravity,
the logarithmic corrections exist in any generic theory of gravity.
Furthermore they are determined purely from the low energy data
-- spectrum of massless fields and their interactions -- and are
insensitive to the spectrum of massive fields and higher derivative
corrections. At the same time, they are not universal since they are
sensitive to what kind of massless fields the theory has, and also
their interactions. Thus
once we compute these logarithmic corrections from the low energy
data, they will provide a testing ground for any proposed ultraviolet
completion of gravity and the description of black hole
microstates in this theory. A microscopic 
theory that does not reproduce the
correct logarithmic corrections must not be the correct
microscopic theory of gravity.

So far the computations have been done for spherically symmetric
black hole solutions and whenever microscopic results are available
in string theory, {\it e.g.} in $\NN=4$ and $\NN=8$ string 
theories in four dimensions\cite{9607026,0412287,0505094,0506249,
0508174,0510147,
0602254,0603066,0605210,0607155,0609109,0506151},
the microscopic and the macroscopic results for the
logarithmic corrections are in perfect agreement. For $\NN=2,3$ and 6
supersymmetric theories in four dimensions there are definite predictions for the
logarithmic corrections from the macroscopic side\cite{1108.3842,1109.0444},
but no microscopic results are available yet.
Some results for non-BPS black holes are also
available\cite{1108.3842}
from the macroscopic side
and yet others have been proposed\cite{1109.0444} but 
there are no known
microscopic results to compare them with.

\begin{table}
\begin{center}\def\st{\vrule height 3ex width 0ex}
\begin{tabular}{|l|l|l|l|l|l|l|l|l|l|l|} \hline 
\qquad \qquad Scaling &  logarithmic correction to the entropy
\st\\[1ex] \hline \hline
$Q_1,Q_5,n\sim \Lambda$, $J\sim \Lambda^{3/2}$
&  $-{1\over 4} (n_V-3) \ln \Lambda$ 
\st\\[1ex] \hline
$Q_1,Q_5,n\sim \Lambda$, $J=0$
  & $-{1\over 4} (n_V+3) \ln \Lambda$ 
\st\\[1ex] \hline
\hline 
\end{tabular}
\caption{Logarithmic corrections to the entropy of a 
BMPV black hole in type IIB string theory compactified
on $K3\times S^1/\ZZZ_\bN$ for $\bN=1,2,3,5,7$. 
$n_V\equiv {48\over \bN+1} + 3$ 
denotes the total number of $U(1)$ gauge fields in the
five dimensional theory. 
The entropy given in this table counts the number of states with
fixed values of the total angular momentum as well as the third 
component of the angular momentum and hence
{\it does not include the degeneracy
factor of $(J +1)$ from the multiplicity of $SU(2)_L$ 
spin $J/2$ states.}  
Each of the results
in this table has been calculated independently in the macroscopic and
microscopic description and in each case we find perfect agreement
between the two sides.
}
\label{t1}
\end{center}
\end{table}

In this paper we extend the computation to extremal rotating black
holes. Before we describe our results it is necessary to say a few
words about the ensemble in which we compute the entropy.
It follows from the general analysis of \cite{0809.3304}
that the extremal black hole computes
the entropy in the microcanonical ensemble, 
carrying fixed charges associated with all the gauge fields on
$AdS_2$. In this case the massless gauge fields
on $AdS_2$ include all the Maxwell fields of the original
theory as well as the gauge fields arising out of the metric due to
rotational 
isometries of the black hole solution. For a rotating black
hole carrying SU(2) angular momentum $J_3=j$, only
the isometry associated with rotation about the third axis
gives rise to massless gauge fields in $AdS_2$ and hence
the entropy computed by quantum entropy function counts states
with fixed $J_3=j$ without any restriction on the total
angular momentum. We shall denote by  $\wt d(j)$ the degeneracy
of states with this restriction. We can also
introduce another ensemble in which we fix the total angular
momentum $\vec J^2$ to $j(j+1)$ besides fixing $J_3$ to $j$.
The corresponding degeneracy will be denoted by $d(j)$. The
latter is the relevant number for $j=0$ since a black hole with
$j=0$ will have full $SU(2)$ isometry and as a consequence
quantum entropy function will count states for which
all components of the angular momentum are fixed to 0.
We note however that $d(j)$ and $\wt d(j)$ are related by the
simple relation $d(j) = \wt d(j+1) - \wt d(j)$; thus knowledge of one
determines the other. 
We shall in fact see that for generic $j$,
$\wt d(j+1)$ is related to $\wt d(j)$ by a multiplicative factor of order
unity (it follows from the fact that entropy scales in the same way
as $j$), and hence $\ln d(j)$ and $\ln \wt d(j)$ differ by an additive
term of order unity and have identical
logarithmic corrections. The situation changes when $j$ takes value
that is parametrically smaller than the generic value; 
this case will be
discussed separately later.

Another remark that is relevant for the supersymmetric black holes
is the relation between index and entropy. Black holes compute degeneracies
whereas the quantity that is robust against changes of parameters and
hence can be compared between the macroscopic and the microscopic
results is an appropriate supersymmetric index. It has however 
been argued
in \cite{0903.1477,1009.3226} that for a supersymmetric black hole 
quantum entropy function also
computes an index and hence we can directly compare the results obtained
from quantum entropy function with the index computed in the
microscopic theory. This argument will be reviewed in \S\ref{sensemble}.

We consider two classes of examples, -- BMPV
black holes in five dimensions\cite{9602065} 
and extremal rotating black holes in four
dimensions. For the former explicit microscopic results are
known\cite{9608096,0605210} and our results are 
in perfect agreement with these
results.
We have shown in
table \ref{t1} the 
logarithmic corrections to the entropy of a 
BMPV black hole in type IIB string theory compactified
on $K3\times S^1/\ZZZ_\bN$ where $\bN$ is a prime integer
(1, 2, 3, 5 or 7), and the $\ZZZ_\bN$ acts as a shift along
$S^1$ and a transformation on $K3$ that preserves 16
supersymmetries. 
For $\bN=1$ this reduces to
the original BMPV black hole in type IIB in 
$K3\times S^1$\cite{9602065} while other values of $\bN$
correspond to BMPV black holes in five dimensional CHL
models\cite{9505054,9506048,9508144,9508154}.
$n_V\equiv {48\over {\bN+1}}+3$ 
denotes the total number of $U(1)$ gauge fields in the
five dimensional theory including any vector field that can come
from dualizing a 2-form field, {\it e.g.} for $\bN=1$
we have $n_V=27$. 
The classical black hole solution under consideration carries
$Q_1$ units of D1-brane charge along $S^1$, $Q_5$ units of
D5-brane charge along $K3\times S^1$, $-n/\bN$ units of
momentum along $S^1$ and $SO(4)=SU(2)_L\times SU(2)_R$
angular momentum $\vec J_R^2=0$, $J_{1L}=J_{2L}=0$,
$J_{3L}={J\over 2}$. 
The table
shows the results for the logarithmic corrections to the entropy 
in the limit when $\Lambda$ is large. 
Each of the results
in this table has been calculated independently in the macroscopic and
microscopic description and in each case we find perfect agreement
between the two sides. For this comparison we need to ensure the
correct choice of ensemble on the microscopic side,
so that the result can be compared with the microcanonical
entropy that the macroscopic side computes. There are two relevant
ensembles: in the first one 
$\vec J_R^2=0$ and $J_{3L}$ is fixed at
$J/2$ and in the second one
$\vec J_R^2=0$, $J_{3L}$ is fixed at
$J/2$ and $\vec J_L^2$ is fixed at ${J\over 2} ({J\over 2}+1)$.
Both ensembles of course have all the charges fixed.
As discussed above,
a direct macroscopic calculation gives the entropy in the first
ensemble for $J\ne 0$ and in the second ensemble for $J=0$.
We have however used the known relation between the entropies in the
two ensembles 
to express all the results in table \ref{t1} in the second ensemble.
Also important for this comparison is the relation
between the index and entropy for black holes preserving four
supersymmetries\cite{0903.1477,1009.3226} so that 
we can sensibly compare the entropy
on the macroscopic side to the logarithm of the index computed
on the microscopic side.
Finally, this analysis can be easily generalized to
other allowed values of $\bN$ and with K3 replaced by 
$T^4$ using the results of \cite{0607155,0609109}, with the 
only difference that the
simple relation between $\bN$ and $n_V$ will be lost, and some
of the modular functions which will appear later in our microscopic
formula will have a more complicated form.

Our results also hold for BMPV black holes in type IIB string
theory compactified on $T^5$. In fact on the macroscopic side there
is no difference between the analysis in the CHL models or
$T^5$ compactifications, except that in the latter case $n_V$ has
a specific value $27$. Thus for the scaling $Q_1, Q_5,n\sim\Lambda$,
$J\sim \Lambda^{3/2}$ the result for logarithmic correction
to the entropy takes the form
\be \label{etypeiitor}
-6\, \ln\Lambda\, .
\ee
On the other hand for $Q_1, Q_5,n\sim\Lambda$, $J=0$ we get
a logarithmic correction of the form:
\be \label{etypeiitorj0}
-{15\over 2}\, \ln\Lambda\, .
\ee
The computation on the microscopic side is quite different since we
have a completely different microscopic 
formula\cite{9903163}. Nevertheless the
final results agree precisely with \refb{etypeiitor} and \refb{etypeiitorj0}.

In computing the entropy of the black hole from the microscopic side
one often invokes the Cardy formula\cite{0005003,0005017}.
Since the underlying CFT has
central charge of order $Q_1Q_5$ and the black hole describes an
ensemble of states
with vanishing $\bar L_0$ eigenvalue and
$L_0$ eigenvalue of order $n$, the central charge scales
as $\Lambda^2$ and the $L_0$ eigenvalue scales as $\Lambda$
in the scaling limit we are
considering.
Thus the Cardy formula is not directly applicable.
But one often uses the intuition that the CFT describing
the D1-D5 system is a sigma model whose target space
is the symmetric product of $Q_1Q_5$ copies of
$K3$ (or $T^4$)\cite{9512078}, 
and the twisted sector of this theory has long string excitations
whose dynamics is described by 
an effective CFT with central charge of order
one and $L_0$ eigenvalue of order 
$Q_1Q_5n\sim \Lambda^3$\cite{9604042,9601152}. 
This effectively amounts to keeping
$Q_1$, $Q_5$ fixed and scaling $n$ as $\Lambda^3$, and
Cardy formula can be applied. We have examined
the microscopic formula to examine the behaviour of the index in this
limit with $J\sim \Lambda^{3/2-\alpha}$ for $\alpha\ge 0$. 
We find that while for type IIB string theory on $T^5$ the result
agrees with the one shown in table \ref{t1} for $n_V=27$, for
the CHL models
(including type IIB string theory on $K3\times S^1$) the logarithmic
correction to the log of the index is given by
$-\alpha\ln\Lambda$,
which 
is quite different from the actual result 
$-{1\over 4} (n_V-3+4\alpha) \ln \Lambda$
(see table \ref{t1} for $\alpha=0$ and
eq.\refb{estretchresult} below for general $\alpha$).
This mismatch for the
CHL models shows that it is not always possible to extract
the logarithmic correction to black hole entropy by examining the
corresponding formula in the Cardy limit.

For non-supersymmetric extreme Kerr black holes
in four dimensions our result for logarithmic correction to the
entropy takes the form:
\be \label{ekerrint}
{1\over 180} \, \ln A_H\, \left( 2 n_S - 26 n_V +7 n_F
- {233\over 2}\, 
n_{3/2} + {64}\right)\, ,
\ee
if, besides gravity, the theory contains $n_S$ minimally coupled
massless scalar, $n_V$ minimally coupled massless U(1) gauge
fields, $n_F$ minimally coupled massless Dirac fermions and
$n_{3/2}$ minimally coupled massless real Rarita-Schwinger
fields.  Here $A_H$ is the area of the event horizon. 
It will be interesting to explore if
\refb{ekerrint} can be explained using
Kerr/CFT correspondence\cite{0809.4266}, which is an
attempt to explain the origin of extremel Kerr black hole entropy by
postulating the existence of a (1+1) dimensional conformal field
theory dual to gravity in the near horizon geometry of the black
hole.

Besides the scaling limits discussed above, we also consider the
cases of slowly rotating black holes -- black holes for which the
angular momentum is parametrically smaller than the other charges
so that the contribution of the angular momentum to the entropy
becomes negligible in the scaling limit. For example for BMPV
black holes we can consider the scaling limit $Q_1, Q_5,n\sim\Lambda$
and $J\sim \Lambda^{{3\over 2}-\alpha}$ for some positive constant
$\alpha$. In this case the computation of the macroscopic entropy
requires some additional assumptions about the spectrum of the
kinetic operator in the background of such slowly rotating black holes.
With this our result for the logarithmic correction to the entropy, in the
ensemble in which $\vec J_R^2=0$, $J_{3L}$ is fixed to be $J/2$ and
$\vec J_L^2$ is fixed to be ${J\over 2}({J\over 2}+1)$, is given by
\be \label{estretchresult}
-{1\over 4} (n_V-3+4\alpha) \ln \Lambda\, .
\ee
This  agrees with the microscopic results in the 
same ensemble. We also consider 
slowly rotating
extremal Kerr-Newmann black hole carrying charge $q$ and
angular momentum $J$, with rotation parameter
$\gamma \equiv J/q^2 \sim J/A_H$ small. In this case 
we get a logarithmic correction
of the form
\be \label{eresrnfinint}
-{1\over 180} (784 + n_S + 62 n_V + 11 n_F) \, \ln {A_H} +\ln{\gamma}\, ,
\ee
if the theory contains, besides the metric and the Maxwell field under
which the black hole is charged, $n_S$ minimally coupled massess scalar
field, $n_V$ minimally coupled {\it additional} massless vector fields and
$n_F$ minimally coupled massless Dirac fermions. 
\refb{eresrnfinint} refers to
the entropy computed  in an ensemble with
fixed charges, $J_{3}=J$ and $\vec J^2 = J(J+1)$.

Various other 
earlier approaches to computing
logarithmic corrections to black hole entropy can be found
in \cite{9407001,9408068,9412161,9604118,
9709064,0002040,0005017,
0104010,0112041,0406044,0409024,0805.2220,
0808.3688,0809.1508,0911.4379,1003.1083,1008.4314}.
Since in \cite{1108.3842}
we have given a detailed comparison between our method
and the
method of 
\cite{9709064,1008.4314,1104.3712} -- which is closest to
the method we use --
we shall not repeat the
discussion here. However it will be prudent to point out that
the naive application of the method of 
\cite{9709064,1008.4314,1104.3712} 
would give zero result for the logarithmic
correction to the BMPV black hole entropy, since the
trace anomaly on which their computation is based vanishes in
odd dimensions. The reason that we get a non-zero result is
due to the zero modes which must be treated separately, and
is not correctly accounted for in the trace anomaly based
approach. For extremal Kerr black hole the analysis of
\cite{9709064,1008.4314,1104.3712} would correctly give the
dependence on $n_S$, $n_V$, $n_F$ and $n_{3/2}$ in
\refb{ekerrint} but would
give the constant term to be 424 instead of 64. 
This is again due to the
additional corrections due to graviton zero modes which have been
included in \refb{ekerrint}.

The rest of the paper is organized as follows. In \S\ref{sstrategy} we
describe the procedure for computing logarithmic correction to the
entropy of an extremal rotating black hole, generalizing our earlier
analysis for spherically symmetric extremal black holes.
In \S\ref{skerr} we apply the results of \S\ref{sstrategy} 
to compute the logarithmic correction to the entropy of an extremal Kerr
black hole in four dimensions, arriving at the result \refb{ekerrint}.
In \S\ref{sbmpv} we repeat the analysis for five dimensional BMPV
black hole in the limit when all the charges are scaled simultaneously
by some large number $\Lambda$, and arrive at the results 
given in table \ref{t1}.
In \S\ref{sbmpvmicro} we carry out the microscopic analysis of the
entropy of BMPV black holes in the same scaling limits
and reproduce the results of table
\ref{t1} and eq.\refb{estretchresult}. 
We find however that while the macroscopic analysis takes
identical form for toroidal compactification of type IIB string theory
and the CHL models, the microscopic analysis for these models are
quite different. Nevertheless at the end the microscopic and the
macroscopic results agree for all models. 
In \S\ref{sslowrot}
we compute logarithmic correction to the entropy of
slowly rotating BMPV black holes in five dimensions and
extremal Kerr-Newmann black hole in four dimensions
and arrive at eqs.\refb{estretchresult} and \refb{eresrnfinint}.
In appendix \ref{sbasis}
we review the counting of zero modes for different fields which play an
important role in the macroscopic computation of logarithmic corrections
to the entropy.

We shall end this section by drawing some general lessons from our
analysis. Since quantum entropy function approach is rooted in
Euclidean quantum gravity\cite{gibbhaw}, what the results of this paper
and those in \cite{1005.3044,1106.0080} show is that 
Euclidean quantum
gravity provides us with an accurate description of
the quantum corrections to black hole entropy. Of course Euclidean 
quantum gravity
path integral suffers from the usual ultraviolet divergences of gravity,
but some computations, {\it e.g.} logarithmic corrections to black hole
entropy, are insensitive to the ultraviolet completion of the theory
and are determined solely by the infrared theory\cite{1104.3712}. 
Thus we can in principle
use Euclidean quantum gravity 
to calculate even the entropy of non-extremal black holes, and any
consistent ultraviolet completion of the theory must reproduce these
results. This could provide
a strong check on any proposed ultraviolet completion of gravity, if the latter
provides an independent derivation of the logarithmic correction to the
black hole entropy.

\sectiono{General strategy} \label{sstrategy}

In \cite{1005.3044,1106.0080,1108.3842} we used 
the quantum entropy function\cite{0809.3304} to compute
logarithmic corrections to the entropy of four dimensional spherically
symmetric extremal black holes with $AdS_2\times S^2$ near horizon
geometry. In this section we shall generalize this analysis to rotating
extremal black holes in arbitrary dimensions.

\subsection{Partition function in the near horizon geometry} \label{sqent}

Rotating extremal black holes are expected to have
an $AdS_2$ factor in their near horizon 
geometry\cite{9905099,0606244,0705.4214,0806.2051,0803.2998},
possibly fibered over some other directions. Besides the
$AdS_2$ factor the near horizon geometry consists of various other
compact directions, some of which could be inherited from the
compact directions of the asymptotic geometry while others could
be the angular coordinates of the non-compact part of the
asymptotic space-time. Our goal will be to compute
logarithmic corrections to the entropy in 
certain scaling limit in which the various charges grow in a
certain way. In this limit some of these compact directions 
grow in size while the others remain fixed in size. 
We shall label by $K$ the space spanned
by the compact directions which grow in size, irrespective of
whether they are inherited from the compact coordinates or the
angular coordinates of the asymptotic space-time.
Let $d$ be the dimension of $K$.
We shall Fourier decompose all string fields along 
the compact directions of fixed size and treat this as a $(d+2)$
dimensional theory with the $(d+2)$ coordinates containing the
coordinates of $AdS_2$ and $K$.
In what follows we shall denote the coordinates of $AdS_2$ by $x$,
the coordinates along $K$ by $y$ and the coordinates $(x,y)$
collectively by $z$.
We shall for simplicity assume that the
radius of curvature of $AdS_2$ and
the sizes of the other large dimensions labelled by $y$ are of the same
order $a$, but this assumption can be easily relaxed. 
The Euclidean near horizon $(d+2)$ dimensional 
metric $g_{\mu\nu}$ can then be written as
\be \label{emets}
g_{\mu\nu} dz^\mu dz^\nu 
= a^2 \, f(y) (d\eta^2 +\sinh^2\eta \, d\theta^2) + a^2 \, ds_K^2\, ,
\ee
where $(\eta,\theta)$ are the coordinates along
$AdS_2$, $f(y)$ is some function of the coordinates $y$
along $K$, and
$ds_K^2$ is constructed out of differentials which are invariant
under the $SL(2,R)$ isometry of $AdS_2$. Note that we have
explicitly factored out the large dimensional parameter $a^2$ so
that $f(y)$ and the metric $ds_K^2$ are of order unity.
Some examples of such
metrics can be found in \cite{0606244} (in a different coordinate
system for labelling $AdS_2$ metric). It follows from the
$SL(2,R)$ isometry of $AdS_2$ that
\be \label{edegg}
\sqrt{\det g} = \sinh \eta \, G(y)\, ,
\ee
for some function $G(y)$ of order $a^{d+2}$. In particular $G(y)$ has no
dependence on the coordinates of $AdS_2$.

Let $Z_{AdS_2}$ denote the partition function of string
theory in the near horizon geometry, evaluated by
carrying out functional integral over all the string fields
weighted by the exponential of the Euclidean action
$\SSS$, 
with boundary conditions such that asymptotically the field
configuration approaches the near horizon geometry
of the black hole.
Since in $AdS_2$ the asymptotic
boundary conditions fix the electric fields, or equivalently
the charges carried by the black hole, and let the
constant modes of the gauge fields fluctuate, we need
to include in the path integral a boundary term
$\exp(-i\ointop \sum_k q_k A^{(k)}_\mu dx^\mu)$ where
$A^{(k)}_\mu$ are the gauge fields and $q_k$ are the
corresponding electric charges carried by the black 
hole\cite{0809.3304}. The list of gauge fields include all the
gauge fields of the original theory, as well as any gauge field
which might arise from dimensional reduction of the metric along
the compact directions of the near horizon geometry.
Thus we have
\be \label{ezads2}
Z_{AdS_2} = \int d\Psi \exp(\SSS
-i\ointop \sum_k q_k A^{(k)}_\mu dx^\mu)\, ,
\ee
where $\Psi$ stands for all the string fields.
$AdS_2/CFT_1$ correspondence
tells us that the full quantum corrected entropy $S_{macro}$
is related to $Z_{AdS_2}$ via\cite{0809.3304}:
\be \label{eads2cft1}
e^{S_{macro} - E_0 L} =  Z_{AdS_2}\, ,
\ee
where $E_0$ is the energy of the ground state
of the black hole carrying a given set of charges,
and $L$ denotes the length of the boundary of $AdS_2$
in a regularization scheme that renders the
volume of $AdS_2$ finite by putting an infrared
cut-off $\eta\le \eta_0$.

Let $\Delta\LL_{eff}(y)$ denote the quantum correction to the $(d+2)$
dimensional effective lagrangian density evaluated in the background
geometry \refb{emets}. Due to the $SL(2,R)$ invariance, $\Delta\LL_{eff}$
is independent of the $AdS_2$ coordinates $x$ but could depend on
the coordinates $y$ of the large compact dimensions.
Then the quantum correction to 
$Z_{AdS_2}$ is given by
\be \label{e2}
\exp\left[
 \int\, \sqrt{\det g} \, d\eta \, d\theta  \, d^d y \, 
 \Delta\LL_{eff}\right]
= \exp\left[2\pi  \, (\cosh\eta_0-1) \, \int d^d y \, G(y)\,
\Delta\LL_{eff}(y)\right]\, .
\ee
The term proportional to
$\cosh\eta_0$ in the exponent
has the interpretation of $-L \Delta E_0
+\OO\left(L^{-1}\right)$
where $L=2\pi \sinh\eta_0$ is the
length of the boundary of $AdS_2$
parametrized by $\theta$ and $\Delta E_0=- \int d^d y \, G(y)\,
\Delta\LL_{eff}(y)$
is the shift in the
ground state energy.\footnote{In order to fix the normalization in the
definition of $L$ and $E$ we need to fix the normalization of the
metric on $AdS_2$ after dimensional reduction on $K$. We can for
example take $(d\eta^2 + \sinh^2\eta d\theta^2)$ as the definition of
the metric on $AdS_2$ in which case the boundary has length
$2\pi \sinh\eta_0$. Any other normalization will simply rescale
$E$ and $L$ in opposite directions without affecting the rest of the
analysis.} Alternatively, this can be cancelled by a boundary
counterterm  which gives a contribution proportional to the length
of the boundary \i.e.\ $\sinh\eta_0$.
The rest of the contribution in the exponent
can be interpreted as the
quantum correction to the black hole 
entropy\cite{0809.3304}. Thus we have
\be \label{e3}
\Delta S_{macro} = -2\pi \, \int d^d y \, G(y)\, \Delta\, \LL_{eff}(y)\, .
\ee
While the term in the exponent proportional to $L$
and hence $\Delta E_0$ can get further corrections
from boundary terms in the action, the finite part
$\Delta S_{macro}$ is defined unambiguously.
This reduces the problem of computing quantum correction to
the black hole entropy to that of computing quantum 
correction to $\LL_{eff}$. If on the other hand we replace
$\Delta\LL_{eff}$ in \refb{e3} by the classical Lagrangian density
then we get back the Wald entropy\cite{0506177} of the
classical black hole.

So far our analysis has been completely general without making any
approximation. However if we are interested in computing the logarithmic
correction to the entropy then we can focus on the contribution to
$\Delta\LL_{eff}$ from the massless fields only\cite{1005.3044,1106.0080}.
We shall now describe the general procedure for calculating one loop
contribution to
$\Delta\LL_{eff}(y)$ from massless fields.

\subsection{Logarithmic correction from the non-zero modes} \label{sorigin}

Let us suppose that the string theory under consideration
contains a set of massless fields $\{\phi_i\}$ (or fields
of mass $\sim a^{-1}$) on $AdS_2\times K$. Here
the index $i$ could run over several scalar fields, or the space-time
indices of tensor fields. Let $f_n^{(i)}(x,y)$ denote an
orthonormal  basis of eigenfunctions of the kinetic operator
expanded around the near horizon geometry, with eigenvalue
$\kappa_n$:
\be \label{eortho}
\int d^2 x\, d^{d} y \, \sqrt{\det g} \, G_{ij} \, f_n^{(i)}(x,y)\,
f_m^{(j)}(x,y) = \delta_{mn}\, ,
\ee
where 
$G_{ij}$ is a metric in the space of fields induced by the
metric on the near horizon geometry, {\it e.g.} for a vector
field $A_\mu$, $G^{\mu\nu}=g^{\mu\nu}$.
 Then the heat kernel $K^{ij}(x,y;x',y')$ is defined as
\be \label{eh1}
K^{ij}(x,y;x',y';s) = \sum_n \, e^{-\kappa_n\, s} \, f_n^{(i)}(x,y)\,
f_n^{(j)}(x',y')\, .
\ee
In \refb{eortho}, \refb{eh1} we have assumed that we are working in a basis
in which the eigenfunctions are real; if this is not the case then we
need to replace one of the $f_n^{(i)}$'s by $f_n^{(i)*}$.
Among the $f_n^{(i)}$'s there may be a special set of
modes for which $\kappa_n$ vanishes. We shall denote these
zero modes by the special symbol $g_\ell^{(i)}(x,y)$, normalized as
\be \label{egnorm}
\int d^2 x\, d^{d} y \, \sqrt{\det g} \, G_{ij} \, g_\ell^{(i)}(x,y)\,
g_{\ell'}^{(j)}(x,y) = \delta_{\ell\ell'}\, ,
\ee
and define
\be \label{ezma}
\bar K^{ij} (x,y;x',y') = \sum_\ell \,  g_\ell^{(i)}(x,y)\,
g_\ell^{(j)}(x',y')\, ,
\ee
\be \label{edefk0}
K(y;s) = G_{ij}\, K^{ij}(x,y;x,y; s)\, , \qquad \bar K(y) = G_{ij}\, 
\bar K^{ij}(x,y;x,y)\, .
\ee
Note that due to the $SL(2,R)$ symmetry, $G_{ij} K^{ij}(x,y;x,y;s)$
and $G_{ij} \bar K^{ij}(x,y;x,y)$ depend only on $y$ but not on $x$.
Using orthonormality of the wave-functions we get
\be \label{eext23}
\int d^2 x \, d^d y \, \sqrt{\det g}\,   
\left(K(y;s) - \bar K(y)\right)={\sum_{n}}' e^{-\kappa_n\, s} \, ,
\ee
where $\sum_n'$ denotes sum over the non-zero modes only.
Since the one loop contribution to $Z_{AdS_2}$ from the non-zero modes
is given by $\prod'_n \kappa_n^{-1/2}=\exp[-{1\over 2}\ln\kappa_n]$,
the contribution to the one loop effective
action can now be expressed as
\be \label{e4}
\Delta\SSS =-{1\over 2}\, {\sum_n}' \ln\kappa_n =
{1\over 2} \int_\eps^\infty {ds\over s} {\sum_n}' e^{-\kappa_n s}
=
{1\over 2}\, \int_{\eps}^\infty\, {ds\over s} \, 
\int d^2 x \, d^d y \, \sqrt{\det g}\,   \left(K(y;s) - \bar K(y)\right)\, ,
\ee
where $\eps$
is an ultraviolet cut-off which we shall take to be of order unity,
\i.e.\ string scale.
Identifying \refb{e4} as the contribution to
$\int d^2 x \, d^d y \, \sqrt{\det g}\, \Delta\LL_{eff}(y)$ we get the contribution
to $\Delta \LL_{eff}(y)$ from the non-zero modes:
\be \label{e3ab}
\Delta\LL_{eff}^{(nz)}(y) = {1\over 2} \, \int_{\eps}^\infty\, {ds\over s} \, 
\left(K(y;s) - \bar K(y)\right)\, .
\ee
Substituting this into \refb{e3} we get the one loop contribution to
$\Delta S_{macro}$ due to the non-zero modes:
\be \label{e3rept}
\Delta S_{macro} = -\pi \, \int_{\eps}^\infty\, {ds\over s} \, 
\int d^d y \, G(y)\, \left(K(y;s) - \bar K(y)\right)\, .
\ee

Since we have labelled by $a^2$ the overall scale factor in the metric,
the non-zero
eigenvalues $\kappa_n$ of the kinetic operator scale as
$1/a^2$ and hence $K(y,s)-\bar K(y)$ 
is a function of $\bar s\equiv s/a^2$.
Thus it is more natural to express \refb{e3rept} as
\be \label{e2.14new}
\Delta S_{macro} = -\pi \, \int_{\eps/a^2}^\infty\, {d\bar s\over \bar s} \, 
\int d^d y \, G(y)\, \left(K(y; s) - \bar K(y)\right)\, .
\ee
In this case
the logarithmic contribution to the entropy -- term proportional to
$\ln a$ -- arises from the $\eps/a^2<<\bar s<<1$, \i.e.\
$\eps<< s << a^2$ region in the $s$
integral.\footnote{If on the other hand the theory has a cosmological
constant, {\it e.g.} for BTZ black holes, then the scale of the
eigenvalues of the kinetic
operator are set by the cosmological constant instead of the parameter
$a$. In this case the entire logarithmic corrections would come only
from integration over the zero modes.}
If we expand $K(y;s)$ in a Laurent series expansion
in $\bar s =s/a^2$ in the region $\bar s << 1$, and if 
$K_0(y)$ denotes
the coefficient of the constant mode in this expansion, then 
using \refb{e2.14new} we see that the
net logarithmic correction to the entropy from the non-zero modes
will be given by
\be \label{enetlognz}
-2 \pi \,  \ln a\, \int d^d y\, G(y)\, \left(K_0(y)  -  \bar K(y)\right)
\, .
\ee

\subsection{Zero mode contribution} \label{szero}

The contribution to $Z_{AdS_2}$ from integration over the zero
modes can be evaluated as follows. First note that we can use \refb{egnorm},
\refb{ezma}, \refb{edefk0} to define the number $N_{zm}$
of zero modes:
\be \label{enz1}
\int d^2 x \, d^d y\, \sqrt{\det g} \, \bar K(y)= \sum_\ell 1 = N_{zm}  
\, .
\ee
In fact often the matrix $\bar K^{ij}$ takes a block diagonal form
in the field space,
with different blocks representing zero modes  of different sets of
fields. In that
case we can use the analog of \refb{enz1} to define the number
of zero modes of each block. If these different blocks are labelled
by $\{A_r\}$ and we define
\be \label{edefkry}
\bar K^r(y) \equiv \sum_{\ell\in A_r} G_{ij}
g_\ell^{(i)}(x,y)\,
g_\ell^{(j)}(x,y)\, ,
\ee
then the number of zero modes belonging to
the $r$-th block will be given by
\be \label{ediffblock}
N^{(r)}_{zm} = \int d^2 x \, d^d y\, \sqrt{\det g}\,   
\bar K^r(y)=2\pi  \,  \left(\cosh\eta_0 - 1
\right) \int d^d y \, G(y) \, \bar K^r(y) \, . \ee
Typically these zero modes are
associated with certain asymptotic symmetries, -- gauge transformation
with parameters which do not vanish at infinity. In this case we
can evaluate the integration over the zero modes by making a change
of variables from the coefficients of the zero modes to the parameters
labelling the (super-)group of asymptotic symmetries.  We shall
use an $a$-independent parametrization of the asymptotic symmetry
group so that the group composition laws and the range over which the
parameters take value is $a$-independent.
Suppose for the zero modes in the $r$'th block the Jacobian for the
change of variables from the fields to supergroup parameters gives a
factor of $a^{\beta_r}$ for each zero mode. 
Then the net $a$ dependent contribution
to $Z_{AdS_2}$ from the zero mode integration will be given by
\be \label{enetzmc}
a^{\sum_r \beta_r N^{(r)}_{zm}}
= \exp\left[2\pi \ln a \, (\cosh\eta_0-1) \, \int \, d^d y \, G(y)\,  
\sum_r \beta_r  
\bar K^r(y)\right]\, .
\ee
As before we can interpret the term in the exponent proportional to
$\cosh\eta_0$ as a contribution to the ground state energy and the finite term as
a contribution to $\Delta S_{macro}$.
Adding this to \refb{enetlognz} we get
\be \label{edeltasbh}
\Delta S_{macro} = -2\pi\, \ln a \, \int d^d y \, G(y)\, 
\left( K_0(y) + \sum_r (\beta_r  -1)
\bar K^r(y)\right)\, .
\ee
We shall refer to the term proportional to 
$\sum_r (\beta_r  -1)\int d^d y \, G(y)\, 
\bar K^r(y)$ as the zero mode contribution although it should
be kept in mind that only the term proportional to
$\beta_r$ arises from integration over the zero modes,
and the $-1$ term is the result of subtracting the zero
mode contribution from the heat kernel to correctly
compute the result of integration over the non-zero modes.

The contribution from the fermionic fields can be included in
the above analysis as follows.
Let $\{\psi^i\}$ denote the set of fermion fields in the theory.
Here $i$ labels the internal indices or space-time vector index
(for the gravitino fields) but the spinor indices are suppressed.
Without any loss of generality we can take the
$\psi^i$'s to be Majorana
spinors satisfying $\bar \psi^i =(\psi^i)^T \wt C$ where $\wt C$
is the charge conjugation operator. Then the kinetic term for the
fermions have the form
\be \label{eferkin}
-{1\over 2} \, \bar\psi^i \DD_{ij} \psi^j = -{1\over 2} \, 
(\psi^i)^T \wt C \DD_{ij} \psi^j\, ,
\ee
for some appropriate operator $\DD$.
We can now proceed to define the heat kernel of the fermions
in terms of eigenvalues of $\DD$ in the usual manner, but  
with the following simple changes. Since the
integration over the fermions produce $(\det \DD)^{1/2}$
instead of $(\det\DD)^{-1/2}$, we need
to include an extra minus sign in the definition of the heat kernel.
Also since the fermionic kinetic operator is linear in derivative,
it will be convenient to first compute the determinant of 
$\DD^2$ 
and then take an additional square root of the determinant. 
This is implemented by including an extra factor 
of $1/2$ in
the definition of the heat kernel.
We shall denote
by $K_0^f(y)$ the $s$-independent part of the trace
of the fermionic heat kernel
in the small $s$ expansion
after taking into account this factor of $-1/2$.
To identify the zero 
modes however
we need to work with the kinetic operator and not its square since
the zero mode structure may get modified upon taking the square
{\it e.g.} the kinetic operator may have
blocks in the Jordan canonical form
which squares to zero,  but the matrix itself may be 
non-zero.\footnote{This problem would not arise if we work
with $\wt C\DD$ instead of $\DD$ since $\wt C \DD$ is 
represented by an
anti-symmetric matrix. However it is easier to work with $\DD^2$
instead of $(\wt C\DD)^2$.}
Let us denote by $\bar K^f(y)$ the total fermion zero mode 
contribution
to $K^f_0(y)$. 
Then we arrive at an expression similar to \refb{enetlognz}
for the fermionic non-zero mode contribution to the entropy:
\be \label{enetlognzfer}
 -2\pi \, \ln a\, \int d^d y \, G(y)\, \left(K^f_0(y) -  \bar K^f(y)\right) \, .
\ee
Next we need to carry out
the integration over the zero modes. 
Taking into account the extra factor of $-1/2$ in the definition
of the fermionic heat kernel we see that the analog of
\refb{ediffblock} for the total number of fermion zero modes 
$N^{(f)}_{zm}$ now
takes the form
\be \label{enfzm}
N^{(f)}_{zm} = -4\pi  \, \left(\cosh\eta_0 - 1
\right)\, \int d^d y \, G(y) \,\bar K^f(y) \, .
\ee
Let us further assume that integration over each fermion zero
modes gives a factor of $a^{-\beta_f/2}$ for some constant
$\beta_f$.
Then the total $a$-dependent contribution from integration
over the fermion zero modes is given by
\be \label{eadep}
a^{-\beta_f N^{(f)}_{zm}/2} =
\exp\left[2\pi \, \ln a\, \left(\cosh\eta_0 - 1
\right)\, \int d^d y\, G(y)\, \beta_f \bar K^f(y)  \right]\, .
\ee
As usual the coefficient of $\cosh\eta_0$ can be interpreted
as due to a shift in the energy $E_0$, whereas the
$\eta_0$ independent term has the interpretation of a
contribution to the black hole entropy.
Combining this with the contribution \refb{enetlognzfer}
from the non-zero modes
we arrive at the 
following expression for the 
logarithmic correction to the entropy from the fermion
zero modes:
\be \label{edeltasbhfermi}
\Delta S_{macro} = -2\pi\, \ln a \, \int d^d y\, G(y)\, 
\left( K^f_0(y) +  (\beta_f  -1)
\bar K^f(y)\right)\, .
\ee
In other words, we can use \refb{edeltasbh} to represent contributions
from both the bosonic and the fermionic modes provided we include the
extra factors of $-1/2$ in the definition of the heat kernel and $\beta_r$ for
the fermions.

At the end of this process we are still left with an $a$-independent
contribution from integration over the supergroup which contains
both bosonic and fermonic zero modes. Using supersymmetric
localization we can get finite result for this integral by
canceling the infinities from the bosonic zero mode integration
against the zeroes from the fermion zero mode 
integration\cite{0905.2686,1012.0265}. However since there is
no $a$-dependence in this contribution we shall not discuss this
any further.

\subsection{Evaluation of $K_0(y)$} \label{sk0y}

In \cite{1005.3044,1106.0080,1108.3842} $K_0(y)$ was evaluated
for various
fields by finding the eigenfunctions and eigenvalues of the kinetic
operator in the $AdS_2\times S^2$ near horizon geometry.
Since this is a difficult task in absence of rotational
symmetry, we shall now describe an
indirect method\cite{duffobs,christ-duff1,christ-duff2,duffnieu,duffroc,
birrel,gilkey,0306138,1009.4439} 
for computing $K_0(y)$ which works under special
circumstances.
It can be argued on general grounds that the small $s$ expansion of
$K(y;s)$ contains even (odd) powers of $s^{1/2}$ in even (odd) dimensions
(see {\it e.g.} \cite{0306138} for a recent review).
As a result
$K_0(y)$, which is the coefficient of the $s^0$ term in the small $s$
expansion of $K(y;s)$, vanishes in odd dimensions. Thus 
in the rest of this subsection
we shall restrict our analysis to the four dimensional
(\i.e.\ $d=2$)  case. In this
case 
one can show 
that\cite{duffobs,christ-duff1,christ-duff2,duffnieu,duffroc,
birrel,gilkey,0306138,1009.4439} 
if we have purely gravitational background, {\it e.g.} as in the case of
extremal Kerr black holes, then
in a theory with $n_S$ minimally
coupled massless scalar fields, $n_V$ minimally coupled
massless vector fields, $n_F$ minimally coupled massless
Dirac fields, $n_{3/2}$ minimally coupled massless spin 3/2 field
and $n_2$ minimally coupled massless spin 2 fields, $K_0(y)$ is
given by
\be \label{ebirrela}
K_0(y) = -{1\over 90\pi^2} (n_S + 62 n_V + 11 n_F) E
- {1\over 30\pi^2} (n_S + 12 n_V + 6 n_F - {233\over 6}
n_{3/2} + {424\over 3} n_2) I\, ,
\ee
where
\ben \label{edefiea}
E &=& {1\over 64} \left(R_{\mu\nu\rho\sigma} 
R^{\mu\nu\rho\sigma} - 4 R_{\mu\nu} R^{\mu\nu}
+ R^2\right) \nn
I &=&  -{1\over 64} \left(R_{\mu\nu\rho\sigma} 
R^{\mu\nu\rho\sigma} - 2 R_{\mu\nu} R^{\mu\nu}
+ {1\over 3} R^2\right)\, .
\een
In pure gravity $R_{\mu\nu}$ and $R$ vanish for classical
solutions, and \refb{ebirrela} can be written as
\be \label{ebirrelb}
K_0(y) = -{1\over 90\pi^2} (-2 n_S + 26 n_V -7 n_F
+ {233\over 2}\, 
n_{3/2} - {424}\, n_2) E
\, .
\ee
In the presence of background electromagnetic field strength
eq.\refb{ebirrelb} gets additional contribution involving powers
of the background field strength. In principle one could write down
the most general four derivative terms on the right hand side of
\refb{ebirrela} and calculate the coefficients of these terms by
perturbative computation of the trace anomaly. Alternatively in
supersymmetric theories one could invoke supersymmetry to
constrain the terms on the right hand side of \refb{ebirrela}. In
particular for
theories with $\NN\ge 2$ supersymmetry in four dimensions one
can argue that \refb{ebirrelb} gives the exact result for
$K_0(y)$\cite{duffroc,1108.3842}. In our analysis
we shall mainly
focus on solutions without any background flux and hence
use \refb{ebirrelb}.

\subsection{Computation of $\beta_r$} \label{sbetar}

We shall now describe the procedure for
computing  $\beta_r$ for various fields
following \cite{1108.3842}.
Let us begin with the contribution from an $U(1)$ gauge field $A_\mu$.
The path integral measure over $A_\mu$ is normalized via
\be \label{eap2bmp}
\int [DA_\mu] \exp\left[- \int d^{d+2} z \, \sqrt{\det g} \, 
g^{\mu\nu} A_\mu A_\nu
\right] = 1\, ,
\ee
where, as mentioned earlier, 
$z^\mu$ stand for both the coordinates 
$x$ along $AdS_2$ and
the coordinates $y$ along the large compact dimensions. 
In the
scaling limit we consider $g_{\mu\nu}$ can be written as 
$a^2 g^{(0)}_{\mu\nu}$ where $a$ scales as some power of 
$\Lambda$ and 
$g^{(0)}_{\mu\nu}$ is an $a$ independent
constant metric.
Thus we can express \refb{eap2bmp} as
\be \label{eap3bmp}
\int [DA_\mu] \exp\left[- a^{d} \int d^{d+2} z \, \sqrt{\det g^{(0)}} \, 
g^{(0)\mu\nu} A_\mu A_\nu
\right] = 1\, .
\ee
{}From this we see that up to an $a$ independent 
normalization
constant, $[DA_\mu]$  actually corresponds to integration
with measure $\prod_{\mu,z} d(a^{d/2} A_\mu(z))$. 
On the other hand  the gauge field
zero modes are associated with
deformations produced by the gauge transformations
with non-normalizable parameters: $\delta A_\mu\propto
\p_\mu \Lambda(z)$ for some functions $\Lambda(z)$ with 
$a$-independent
integration range. Thus
the result of integration over the gauge field zero
modes can be found by first changing the integration over the
zero modes of
$(a^{d/2} A_\mu)$ to integration over $\Lambda$ and then
picking up the contribution from the Jacobian in this
change of variables. This gives a factor of $a^{d/2}$ 
from integration
over each zero mode of $A_\mu$.
Comparing this with the definition of $\beta_r$ given above
\refb{enetzmc} we see that for gauge fields we have $\beta_v=d/2$.

The effect of integration over the zero modes of the 
fluctuations
$h_{\mu\nu}$ of the metric (including those of the
gauge fields arising from the dimensional
reduction of the metric along the large compact dimensions) 
can be found in a similar way,
with \refb{eap2bmp}, \refb{eap3bmp} replaced by
\be \label{ebp2bmp}
\int [Dh_{\mu\nu}] \exp\left[- \int d^{d+2} z \, \sqrt{\det g} \, 
g^{\mu\nu} g^{\rho\sigma} h_{\mu\rho} h_{\nu\sigma}
\right] = 1\, ,
\ee
\i.e.\
\be \label{ebp3bmp}
\int [Dh_{\mu\nu}] \exp\left[- a^{d-2} \int d^{d+2} z \, 
\sqrt{\det g^{(0)}} \, 
g^{(0)\mu\nu} g^{(0)\rho\sigma}h_{\mu\rho} h_{\nu\sigma}
\right] = 1\, .
\ee
Thus the correctly normalized integration
measure, up to an $a$ independent constant, is 
$\prod_{z,(\mu\nu)} d(a^{(d-2)/2}h_{\mu\nu}(z))$.
Now the zero modes of the metric are associated
with diffeomorphisms with non-normalizable parameters:
$h_{\mu\nu}\propto D_\mu\xi_\nu + D_\nu\xi_\mu$, with
the diffeomorphism parameter $\xi^\mu(z)$
having $a$ independent integration range. Thus the $a$
dependence of the integral over the metric zero modes
can be found by finding the Jacobian from the change
of variables from $a^{(d-2)/2}h_{\mu\nu}$ to $\xi^\mu$. Lowering
of the index of $\xi^\mu$ gives a factor of $a^2$, leading to
a factor of $a^{(d+2)/2}$ per zero mode. Thus for the metric we have
$\beta_m=(d+2)/2$.

Next we turn to the contribution due to the zero modes of
the gravitino field $\psi_\mu$. In this case eqs.\refb{eap2bmp}, 
\refb{eap3bmp}
are replaced by:
\be \label{eap2bmpfer}
\int [D\psi_\mu] \exp\left[- \int d^{d+2} z \, \sqrt{\det g} \, g^{\mu\nu} \bar\psi_\mu 
\bar\psi_\nu
\right] = 1\, ,
\ee
\i.e.\
\be \label{eap3bmpfer}
\int [D\psi_\mu] \exp\left[- a^{d} \int d^{d+2} z \, \sqrt{\det g^{(0)}} \, 
g^{(0)\mu\nu} \bar\psi_\mu \bar\psi_\nu
\right] = 1\, .
\ee
Thus up to an $a$ independent normalization constant $[D\psi_\mu]$
stands for $\prod d(a^{d/2}\psi_\mu)$. On the other hand these
zero modes are associated with the
deformations corresponding to local supersymmetry
transformation ($\delta\psi_\mu\propto D_\mu\eps$) with
supersymmetry transformation parameters $\eps$ which
do not vanish at infinity. Now since the anti-commutator of
two supersymmetry transformations correspond to a
general coordinate transformation with parameter
$\xi^\mu =\bar \eps \gamma^\mu \eps$, and since
$\gamma^\mu\sim a^{-1}$, we conclude that
$\eps_0=a^{-1/2}\eps$
provides a parametrization of the asymptotic supergroup
in which the group composition laws become $a$-independent. 
Writing $\delta(a^{d/2}\psi_\mu)
\propto a^{(d+1)/2}D_\mu\eps_0$, using the fact that the
integration over the supergroup parameter $\eps_0$ produces
an $a$ independent result and that $d(\lambda \eps_0) =
\lambda^{-1} d\eps_0$ for a grassmann  variable $\eps_0$, 
we now see that each
fermion zero mode integration produces a factor of
$a^{-(d+1)/2}$. 
Comparing this with the definition of $\beta_f$ given below
\refb{enfzm} we see that $\beta_f=d+1$. 

The results of this subsection can be
summarized in the relations
\be \label{ebetarexp}
\beta_v = {d\over 2}, \qquad \beta_m = {d+2\over 2}, \qquad
\beta_f = d+1\, .
\ee
For $d=2$ this reproduces the results of \cite{1108.3842}.

\subsection{Computation of $\int d^d y \, G(y)\, 
\bar K^r(y)$} \label{snumzero}

Finally we shall discuss the
computation of $\int d^d y \, G(y)\, 
\bar K^r(y)$ for various
fields. This is best done with the help of \refb{ediffblock},
\refb{enfzm} and 
the results in appendix \ref{sbasis}.
For each massless gauge field on $AdS_2$
the number of zero modes is given by
eq.\refb{etotalvz} and for each massless symmetric tensor field 
on $AdS_2$ the number
of zero modes is given by \refb{ediscretemetric}. 
Comparing these with \refb{ediffblock} we see that each gauge field on
$AdS_2$ contributes a factor of $1$ to $2\pi\, \int d^d y \, G(y)\, 
\bar K^r(y)$ and each symmetric rank two tensor field in $AdS_2$
contributes a factor of $3$ to $2\pi\, \int d^d y \, G(y)\, 
\bar K^r(y)$. Thus for example if we consider the extremal Kerr
black hole in four dimensions whose near horizon geometry has a
squashed sphere with U(1) symmetry besides the $AdS_2$ factor,
we get a U(1) gauge field and a symmetric rank two tensor field
on $AdS_2$ after we dimensionally reduce the metric on the squashed
sphere. This gives a net contribution of $1+3=4$ to
$2\pi\, \int d^d y \, G(y)\, 
\bar K^r(y)$. Finally for computing $\int d^d y \, G(y)\, 
\bar K^f(y)$ for BMPV black holes
we can use the result \refb{etotfin} for the total number of
fermion zero modes. 
Identifying this with \refb{enfzm} we get
$-4\pi\, \int d^d y \, G(y)\, 
\bar K^f(y)=8$.
We can test this for quarter BPS black holes in four
dimensional $\NN=4$
supersymmetric string theories or half BPS black holes in four
dimensional $\NN=2$
supersymmetric string theories which have  
identical number of fermion zero modes in the
near horizon geometry. 
For these cases we have $d=2$, 
$\beta_f=3$ and
the net contribution of the fermion zero modes
to $\Delta S_{macro}$ 
computed from \refb{edeltasbhfermi} will be
$8\ln a$. This agrees with the results of 
\cite{1106.0080,1108.3842}.

\sectiono{Extremal Kerr black hole in four dimensions}  \label{skerr}

We now turn to the analysis of logarithmic corrections to the entropy
of an extremal Kerr black hole solution in Einstein gravity.
We take the gravitational part of the action to be
\be\label{eaction}
\SSS=\int d^4 x \sqrt{-\det g}\, \LL, \qquad \LL=R\, .
\ee
We shall assume that the theory has, besides the metric, $n_S$
minimally coupled massless scalars, $n_V$ minimally coupled
massless vector fields, $n_F$ minimally coupled massless Dirac
fermions and $n_{3/2}$ minimally coupled massless Majorana
Rarita-Schwinger fields.
The
near horizon geometry of an extremal Kerr black hole in this
theory is given by (see {\it e.g.} \cite{9905099,0606244} 
where the near horizon
solution is written down in somewhat different coordinate systems)
\be \label{ekerrnear}
ds^2 = a^2\,
(1+\cos^2\psi) \left\{- (r^2-1) dt^2 + dr^2 / (r^2-1) +
d\psi^2\right\} + 4a^2\, {\sin^2\psi
\over 1+\cos^2\psi} (d\phi- (r-1) dt)^2
\ee
where $(\phi,\psi)$ label the azimuthal and the polar coordinates,
$(r,t)$ denote the radial and the time coordinates
and $a$ is a constant related to the angular momentum $J$ via
the relation
\be \label{ejexp}
J = 16\pi a^2\, .
\ee
The classical Bekenstein-Hawking entropy of the black hole,
obtained as $1/4G_N = 4\pi$ times the area of the event horizon, 
is given by
\be\label{entkn}
S_{BH} = 8\pi a^2\int_0^{2\pi} d\phi \int_0^\pi \sin\psi d\psi
= 32\pi^2 a^2 = 2\pi J\, .
\ee

In order to compute the logarithmic correction to the entropy we
first write down the Euclidean near horizon geometry by
replacing $t$ by $-i\theta$. We also introduce the new radial
coordinate $\eta = \cosh^{-1}r$ for convenience.
This gives
\be \label{ekerrneareuclid}
ds^2 = a^2\,
(1+\cos^2\psi) (d\eta^2 + \sinh^2\eta d\theta^2 +
d\psi^2) + 4 a^2\, {\sin^2\psi
\over 1+\cos^2\psi} (d\phi+ i (\cosh\eta-1) d\theta)^2\, .
\ee
Substituting this into eq.\refb{edegg} we get
\be \label{egyexp}
G(y) = 2 \, a^4 \, \sin\psi \, (1+\cos^2\psi)\, .
\ee
Using \refb{edeltasbh}, \refb{edeltasbhfermi} and carrying out the
$\phi$ integral we now get
\be \label{efinkerrexp}
\Delta S_{macro} = -8\pi^2 a^4\, \ln a \, \int d\psi \, \sin\psi\, 
(1+\cos^2\psi)\,
\left( K_0(\psi) + \sum_r (\beta_r  -1)
\bar K^r(\psi)\right)\, ,
\ee
where the sum over $r$ runs over the bosonic as well as the
fermionic fields.

We can now use \refb{ebirrelb}
for computing $K_0(\psi)$.
In this case
$R_{\mu\nu}=0$, $R=0$ and\cite{9912320,0302095}
\be \label{eriemann}
R_{\mu\nu\rho\sigma} R^{\mu\nu\rho\sigma}
= {48\over a^4} {1\over (1+\cos^2\psi)^6} (1 - 15\cos^2\psi +
15\cos^4\psi -\cos^6\psi)\, .
\ee
This gives
\ben \label{ek0kerr}
K_0(\psi) &=& 
{1\over 120\pi^2 a^4 } \left( 2 n_S - 26 n_V +7 n_F
- {233\over 2}\, 
n_{3/2} + {424}\right)\nn && \times
 {1\over (1+\cos^2\psi)^6} 
(1 - 15\cos^2\psi +
15\cos^4\psi -\cos^6\psi)\, .
\een
Substituting this into \refb{efinkerrexp} and using
the result
\be \label{eintegral}
\int_0^\pi \, d\psi\, \sin\psi \, (1+ \cos^2\psi)^{-5}
(1 - 15\cos^2\psi +
15\cos^4\psi -\cos^6\psi) = -{1\over 6}\, ,
\ee
we get the non-zero mode contribution to the 
black hole entropy
to be:
\be \label{ektotkerr}
 -8\pi^2 a^4\, \ln a \, \int d\psi \, \sin\psi\, 
(1+\cos^2\psi)\,
 K_0(\psi) =
 {1\over 90} \, \ln a\, \left( 2 n_S - 26 n_V +7 n_F
- {233\over 2}\, 
n_{3/2} + {424}\right)\, .
\ee
This coincides with the results in \cite{9709064,1008.4314,1104.3712}.

The contribution 
from the zero modes can be computed as follows. 
Possible zero modes in this case arise from the gauge fields
and the metric, -- since the black hole is non-supersymmetric there
are no gravitino zero modes. Since here $d=2$,
\refb{ebetarexp} gives 
$\beta_v=1$ and $\beta_m=2$\cite{1108.3842}.
Eq.\refb{edeltasbh} now shows that the contribution from
the gauge field zero modes, being proportional to $(\beta_v-1)$,
vanishes. To find the contribution from the metric zero modes we
note that since the near horizon geometry has U(1) rotational
symmetry, the dimensional reduction of the metric along the
$(\psi,\phi)$ direction gives a $U(1)$ gauge field and a massless
symmetric rank 2 tensor field on $AdS_2$. According to the
discussion in  \S\ref{snumzero} we get a
net contribution of $(1+3)=4$ to $2\pi \int d^d y \, G(y)\, \bar K^m(y)$.
The second term in eq.\refb{edeltasbh} now 
gives  a net contribution of $-4\ln a$ to $S_{macro}$. Adding this
to the non-zero mode contribution \refb{ektotkerr} we get the
net contribution to $\Delta S_{macro}$ to be
\ben \label{efinkerr}
&& {1\over 90} \, \ln a\, \left( 2 n_S - 26 n_V +7 n_F
- {233\over 2}\, 
n_{3/2} + {64}\right) \nn
&=&
{1\over 180} \, \ln A_H\, \left( 2 n_S - 26 n_V +7 n_F
- {233\over 2}\, 
n_{3/2} + {64}\right) 
\, ,
\een
where $A_H \propto a^2$ is the area of the event horizon.

Let us denote by $C_{tot}$ the net coefficient of the $\ln A_H$
term appearing in \refb{efinkerr}. 
In order to seek a microscopic explanation of this
result we need to specify for which ensemble 
\refb{efinkerr} gives the logarithmic correction to the
entropy. As discussed in the introduction, this gives
logarithmic correction to $\ln\wt d(J)$, $\wt d(J)$ being
number that 
counts all states with fixed $J_3=J$ and all gauge charges
set to 0, but no restriction on
$\vec J^2$. 
We can extract from this the number $d(J)$ where
$\vec J^2$ is also fixed to be $J(J+1)$ as follows.
First note that 
while $\wt d(J)$ counts all states with fixed 
$J_{3}=J$ and
$\vec J^2 \ge {J} \left({J}+1\right)$,
$\wt d(J+1)$ counts all states with fixed $J_{3}$ and
$\vec J^2 \ge \left({J}+1\right) \left({J}+2\right)$.
Furthermore due to $SU(2)$ symmetry, once we fix $\vec J^2$,
the index is independent of the chosen value of $J_{3}$, and hence in
both cases we can take $J_{3}=J$.
Thus $\wt d(J)-
\wt d(J+1)$ will
count all states with $\vec J^2 = {J} \left({J}+1\right)$
and $J_{3}=J$. 
This is the desired microscopic index $d(J)$. Thus we have
\be \label{edjdjt}
d(J) = \wt d(J+1) - \wt d(J) 
= e^{S_{BH}(J+1) + C_{tot}\ln A_H +\cdots } - e^{S_{BH}(J) 
+ C_{tot}\ln A_H +\cdots } \, ,
\ee
where $S_{BH}(J)= 2\pi J$ is the classical Bekenstein-Hawking entropy
given in \refb{entkn} and $\cdots$ denote terms of order unity. It
is easy to see using \refb{entkn}
that the right hand side is given by
$\left(e^{2\pi} -1\right)$ times $\wt d(J)$. Thus the logarithmic correction
to $\ln d(J)$ is the same as that for $\ln \wt d(J)$, and is given by
\refb{efinkerr}.

It will be interesting to explore if Kerr/CFT 
correspondence\cite{0809.4266} 
or any
other approach could give us a microscopic explanation of
\refb{efinkerr}.

\sectiono{BMPV black holes}  \label{sbmpv}

In this section we shall analyze logarithmic corrections 
to the entropy
of a five dimensional BMPV black hole in type IIB string theory
on $K3\times S^1$, as well as in a class of 
CHL models\cite{9505054,9506048,9508144,9508154}
obtained by taking $\ZZZ_\bN$ orbifolds of type IIB string theory
on $K3\times S^1$. The 
$\ZZZ_\bN$ transformation
acts by $2\pi/\bN$ shift along
$S^1$ and an appropriate action on $K3$ that preserves 16 
supersymmetries. 
Our analysis also holds for BMPV black holes in type IIB
string theory on $T^5$ and its various orbifolds discussed in
\cite{0607155,0609109} with the only change that the metric
$\wh g_{mn}$ in \refb{ep9} represents metric on $T^4$.
This macroscopic 
result for logarithmic correction to the
entropy will then be compared with the
microscopic results derived in \S\ref{sbmpvmicro}.

Some recent discussion on higher derivative corrections to
five dimensional black hole entropy can be found in
\cite{0702072,0703087,0705.1847,0801.1863,
0807.0237,0910.4907,0912.0030,1001.1452,0710.3886,
0809.4954,1009.3226}. The results of these
papers imply agreement between macroscopic and microscopic
entropy at the leading order (charge$^{3/2}$) and the first
subleading order (charge$^{1/2}$). The logarithmic corrections
to be studied here constitute the next leading correction to the
entropy.

\subsection{The near horizon geometry} \label{snearbmpv}

In the $\alpha'=1$ unit the ten dimensional action of type IIB
string theory takes the form
\ben\label{tendaction}
S &=& \int d^{10} x \, \sqrt{-\det G} \, \LL,  \nonumber \\
\LL &=&
{1 \over (2\pi)^7}  \left[
                e^{-2 \Phi}\left(R + 4 G^{MN} \p_M\Phi \p_N \Phi\right)
                 - {1 \over12} F^{}_{MNP} F^{MNP} \right],
\een
where $G_{MN}$ is the ten dimensional string metric, $\Phi$ is the
dilaton and $F_{MNP}^{}$ is the RR 3-form field strength under which
the D1 and D5 branes are electrically and magnetically charged.
We shall choose the coordinate along $S^1$ such that $S^1/\ZZZ_\bN$
has period $2\pi R_5$, and as we move $2\pi R_5$ along this direction we
come back to the same point on the circle but a $\ZZZ_\bN$ transformed
point on $K3$. A unit momentum will be defined as $1/R_5$.
Since there are four non-compact space
directions, this theory has an $SO(4)=SU(2)_L\times SU(2)_R$ 
rotational symmetry. We shall denote by $J_{iL}$ and $J_{iR}$ for
$i=1,2,3$ the generators of the $SU(2)_L$ and $SU(2)_R$ rotation
groups.
In this theory we consider a classical BPS black hole solution
carrying
$Q_5$ units of D5-brane charge along $K3\times S^1$, $Q_1$
units of D1-brane charge along $S^1$ (including the $-Q_5$ units
of D1-brane charge that is induced by wrapping $Q_5$ D5-branes on
$K3$), $-n/\bN=-\wt n$ 
units of momentum along
$S^1/\ZZZ_\bN$, $J/2$ units of $J_{3L}$ charge and 
zero $J_{1L}$, $J_{2L}$ and $J_{iR}$ charge\cite{9602065}.
The Lorentzian ten dimensional
near horizon geometry takes the form\cite{0611143,0901.0359}
\ben \label{ep9}
dS^2&=&r_0 {d\rho^2\over \rho^2} + d\chi^2 
+r_0(dx^4 + \cos\psi d\phi)^2
+{\wJ \over 4 r_0} d\chi (dx^4 + \cos\psi d\phi) 
- 2\rho d\chi d\tau
 \nonumber \\ &&  +r_0  \left(
 d\psi^2 + \sin^2\psi d\phi^2\right) 
+ \wh g_{mn} du^m du^n\, , \nonumber \\
e^{\Phi} &=& \lambda\, , \nonumber \\
F^{} &=& {r_0\over \lambda}\, \left[\eps_3 + *\eps_3
+{\wJ \over 8\, r_0^2}\, d\chi\wedge \left({1\over \rho}\, d\rho\wedge
(d \, x^4 +\cos\psi\, d\phi)
+ \sin\psi\, d\psi\wedge d\phi\right)\right]\, ,
\nn &&  (\psi,\phi, x^4)\equiv (2\pi-\psi, \phi+\pi, x^4+\pi)
\equiv (\psi, \phi+2\pi,x^4+2\pi) \equiv (\psi,\phi,x^4+4\pi)\, ,\nn
&& (\chi, \vec u) \equiv (\chi+2\pi R_5, h\, \vec u)\, .
\een
Here
$u_m$ are coordinates along $K3$, $\wh g_{mn}$ is the metric
along $K3$, $h$ represents the action of the $\ZZZ_\bN$ generator
on the coordinates of $K3$, $\lambda$ is an arbitrary constant,
$\eps_3=\sin\psi \, dx^4\wedge d\psi\wedge d\phi$ 
denotes the volume form on a 3-sphere of coordinate radius 2 labelled
by the coordinates $(x^4,\psi,\phi)$
and $*$ denotes Hodge dual in six dimensions spanned by the
coordinates $t$, $\chi$, $\rho$, $x^4$, $\psi$ and $\phi$. The constants
$r_0$, $R_5$, $\wt J$ and the volume $V$ 
of $K3$ are determined in terms
of the charges via the relations
\be \label{edefpar}
r_0={\lambda Q_5\over 4}, \quad R_5 = \sqrt
{\lambda \wt n\over Q_1}, 
\quad \wt J = {J\over 2} \, {Q_5\over \sqrt{Q_1
\wt n}}\,
\lambda^{3/2}, \quad V\equiv \int d^4 u \, \sqrt{\det \wh g} 
= (2\pi)^4 \, {Q_1\over Q_5} \, .
\ee
The Bekenstein-Hawking entropy of this
black hole can be computed by
dividing the area of the event horizon,
spanned by the coordinates $(\chi,x^4,\theta, \phi, \vec u)$,
by four times the effective Newton's constant at the horizon, read
out from \refb{tendaction}-\refb{edefpar}. The result is:
\be \label{eshbbmpv}
S_{BH} =2\pi \sqrt{Q_1 Q_5 \wt n - {J^2 \over 4}}\, .
\ee

In order to make the $SL(2,R)$ symmetry of $AdS_2$ manifest,
we define
\be \label{echangec}
A=\sqrt{r_0} \, 
\left(1 - {\wt J^2 \over 64 r_0^3}\right)^{-1/2}, \qquad
B = -{\wt J\over 8 r_0^2} \, A \,,
\ee
and change coordinates to
\be \label{echangeb}
\wt \tau = A \tau/r_0\, .
\ee
In these coordinates the near horizon metric takes the form
\ben \label{echanged}
dS^2&=&r_0 {d\rho^2\over \rho^2} - r_0\rho^2 d\wt\tau^2
+ \left(d\chi - A
\rho d\wt\tau\right)^2
+r_0(dx^4 + \cos\psi d\phi - B \rho \, d\wt\tau)^2
+r_0  \left(
 d\psi^2 + \sin^2\psi d\phi^2\right) 
\nonumber \\ && 
+{\wJ \over 4 r_0} (d\chi-A\rho d\wt\tau) (dx^4 + \cos\psi d\phi-B
\rho d\wt\tau)
  + \wh g_{mn} du^m du^n\, .
\een
This metric has an
$SL(2,R)$ isometry generated by\cite{9905099}
\be \label{esl2r}
L_1=\partial_{\wt\tau}, \qquad
L_0={\wt\tau}\partial_{\wt\tau}-\rho\partial_\rho, \qquad
L_{-1}=(1/2)(1/\rho^2+\wt\tau^2)\partial_{\wt\tau} - ({\wt\tau}\rho)
\partial_\rho + (A/\rho)
\partial_\chi + (B/\rho) \p_{x^4}\, .
\ee
In order to make connection with the coordinate systems used in
\refb{emets} we need to make a coordinate change
\be \label{etautoeta}
\cosh\eta = {1\over 2} (\rho +\rho^{-1} - \rho\wt\tau^2), \qquad 
e^{-2t} = {(1-\wt\tau)^2 -\rho^{-2}\over (1+\wt\tau)^2 -\rho^{-2}}\, ,
\ee
together with some $(\rho,\wt\tau)$ dependent
shifts on the coordinates $\chi$ and $x^4$ and then
make the analytic continuation $t\to -i\theta$. The classical entropy
function method\cite{0506177}, 
applied to the near horizon geometry with the
action \refb{tendaction}, give the same result for the entropy as
\refb{eshbbmpv} as expected\cite{0611143}.

The scaling limit we shall consider is 
\be \label{escalinglimit}
Q_1 \sim \Lambda, \quad Q_5\sim \Lambda, \quad \wt n \sim \Lambda,
\quad J\sim \Lambda^{3/2}\, ,
\ee
with $\Lambda$ large. 
Also we shall keep the undetermined constant $\lambda$ fixed as
we take the large charge limit.
In this limit $r_0$
grows as $\Lambda$, $\wt J$ grows as $\Lambda^{3/2}$, 
$R_5$ 
and the volume $V$ of $K3$ remains fixed,
and the size $a$ of 
$AdS_2$ spanned by $(\rho,\wt\tau)$ as well as the
the size of the squashed 3-sphere labelled by $(x^4,\psi,\phi)$ grow
as
\be \label{ealambda}
a\equiv \sqrt{r_0}\sim \Lambda^{1/2}\, .
\ee 
Thus in this situation we can apply the formalism
developed in \S\ref{sstrategy} for computing logarithmic correction to the
entropy. For this we dimensionally reduce the theory on the five
directions spanned by the coordinates $(\vec u, \chi)$ and regard this
as a five dimensional theory living on the space spanned by the
coordinates $(x^4,\psi,\phi,\rho,\wt\tau)$. We also dualize
the 3-form field strength $F_{MNP}$ as well as all other 3-form field
strengths to 2-form field strengths so that we can apply the formalism
of \S\ref{sstrategy}.

\subsection{Ensemble choice and index from entropy} 
\label{sensemble}

In this subsection we shall make some comments on
the ensemble in which we compute the entropy. As argued in
\cite{0809.3304}, the quantum entropy function computes 
the entropy in the
microcanonical ensemble in which all charges and angular momenta,
which have the interpretation of charges associated with gauge fields
in $AdS_2$,
are fixed. This means in particular that for the BMPV black hole,
$J_{iR}=0$ and $J_{3L}=J/2$. 
This is exactly analogous to the situation for extremal Kerr black hole
as discussed at the end of \S\ref{skerr}.
On the other hand for the Strominger Vafa black hole carrying zero
angular momentum all components of $\vec J_L$ are associated with
gauge charges on $AdS_2$ and hence we have $J_{iR}=0$ and
$J_{iL}=0$.

Now while comparing our result with the microscopic results we need
to use a protected index instead of the 
degeneracy\cite{9204102,9608096,9903163,0901.0359,1009.3226}. 
In the present situation
we can consider two different indices. One of them will be defined as 
\be \label{edefind}
d(n,Q_1,Q_5,J)\equiv
-{1\over p!}\,
Tr\left[ (-1)^{2 J_{3R}} \, (2 J_{3R})^p
\right]\, ,
\ee
where $p$ takes the value 2 for type IIB string theory on $K3\times S^1$
and the CHL models, but is 6 for type IIB string theory on 
$T^5$.\footnote{This 
agrees with that used in \cite{0901.0359,1009.3226} but 
apparently differs from those in the earlier papers {\it e.g.}
\cite{9608096,9903163}. 
For example for type IIB on $K3\times S^1$, \cite{9608096} would have
$p=0$ whereas for type IIB on $T^5$, \cite{9903163} would have
$p=2$. 
This difference can be attributed to the fact that 
the trace in \refb{edefind} is taken 
over all the modes of the system, whereas in
the definition of the index given in \cite{9608096,9903163} the trace
over the D1-D5 center of mass modes was factored out.
The definition we are using is in the same spirit as the helicity trace index
used in \cite{9611205,9708062} for four dimensional black holes.}
The trace is taken over all states carrying fixed $Q_1$, 
$Q_5$, $n$
and $J_{3L}=J/2$, $\vec J_L^2={J\over 2} \left({J\over 2}+1\right)$  but
different values of $J_{3R}$ and $\vec J_R^2$.
The second index will be defined as
\be \label{edefindtilde}
\wt d(n,Q_1,Q_5,J)\equiv
-{1\over p!}\,
\wt{Tr}\left[ (-1)^{2 J_{3R}} \, (2 J_{3R})^p
\right]\, ,
\ee
where $\wt{Tr}$ denotes that while taking the trace we sum over all
states with fixed $Q_1$, 
$Q_5$, $n$
and $J_{3L}=J/2$,  but
different values of $J_{3R}$, $\vec J_R^2$ and $\vec J_L^2$.
The difference between the two indices is that the $\vec J_L^2$
value is not fixed in the second index.
The two indices are related by a formula similar to \refb{edjdjt}:
\be \label{edjdiff}
d(n,Q_1, Q_5, J)
= \wt d(n,Q_1, Q_5, J) - \wt d(n,Q_1, Q_5, J+2)\, .
\ee

It has been argued in detail
in \cite{0809.3304,1009.3226} that on the 
black hole side the exponential of the entropy
in fact computes an index. This essentially follows from 
factoring the trace in \refb{edefind} or
\refb{edefindtilde} into a trace over the horizon 
degrees
of freedom and the trace over the hair modes --  
modes living outside the horizon. 
In particular the hair modes include
a set of $2p$
fermion zero modes associated with broken supersymmetry which are
charged under $\vec J_R$ but not under $\vec J_L$. 
$Tr[(-1)^{2J_{3R}} (2J_{3R})^p]$ appearing in \refb{edefind}
receives a non-zero constant contribution from these modes, but each
of the $p$ factors of $2J_{3R}$ are needed to prevent the
contribution from vanishing.
After factoring out trace over these zero modes, we are left with
$Tr(-1)^{2J_{3R}}$ with the trace taken over the rest of the modes. 
Since
BPS black holes with four unbroken supersymmetries
are forced to have $\vec J_R=0$\cite{1009.3226}, 
all the states represented by the black hole horizon have 
$(-1)^{2 J_{3R}}=1$, and hence the  contribution to the index
from the degrees of freedom associated with the
horizon will be given by $Tr(1)=e^{S_{macro}}$.
If there are no additional hair modes besides the zero modes
mentioned above then this allows us to compare 
$\exp[S_{macro}]$ directly with the
microscopic index. 
A similar argument holds for the index \refb{edefindtilde}.
Whether one also fixes $\vec J_L^2$ or not
depends on the situation; for non-zero $J$, the black hole
counts states with fixed $J_{3L}=J/2$ but all values of
$\vec J_L^2$. Thus $\wt d$ is the relevant index. On the other
hand for $J=0$, all components of $\vec J_L$ are gauge charges
and fixed to be zero. Thus the index $d$  is the
correct choice.

If there are additional hair modes then we
also need to compute their contribution to the index and
combine this with the contribution $\exp[S_{macro}]$ from the horizon
modes.
It is however more convenient to
identify the macroscopic contribution to the index
as the contribution 
\be \label{emacrobmpv}
\exp[S_{macro}(n,Q_1,Q_5,J)]\, ,
\ee
from the horizon modes only
and {\it remove the
contribution of the (macroscopic) hair modes from the microscopic index before
comparing the macroscopic and the microscopic results}.
It will be argued
in \S\ref{saddhair} that in the scaling limit that we are
considering here, removal of the hair modes from the microscopic
index does not change the coefficient of the logarithmic correction to the
entropy and hence \refb{emacrobmpv} can be directly compared to the
microscopic result for the logarithmic corrections.

\subsection{Logarithmic corrections} \label{slogbmpv}

We now describe the macroscopic computation of logarithmic
corrections to the BMPV black hole entropy.  
According to \refb{edeltasbh} this involves three parts: computation
of $K_0(y)$, computation of $\beta_r$ and computation of $\bar K^r(y)$.
As already discussed in \S\ref{sk0y}, $K_0(y)$ vanishes in odd
dimensions. Thus 
we are left with
only the contribution from the second term of \refb{edeltasbh}.

Let us begin with the contribution from a $U(1)$ gauge field 
$A_\mu$.
Eq.\refb{ebetarexp} gives $\beta_v=3/2$ for $d=3$.
On the other hand since each five dimensional gauge field upon
dimensional reduction on the squashed $S^3$ in the near horizon
geometry gives a gauge field on $AdS_2$, we have, from 
\S\ref{snumzero}, that
$2\pi\, \int d^d y \, G(y)\, 
\bar K^v(y)=1$. Eq.\refb{edeltasbh} now shows that for each
gauge field we have a contribution of $-{1\over 2} \ln a$ to
$\Delta S_{macro}$. Since $a\sim \Lambda^{1/2}$ we see that
$n_V$ vector fields will
give a logarithmic correction of the form
$-(n_V/4) \ln\Lambda$.

The effect of integration over the zero modes of the 
fluctuations
$h_{\mu\nu}$ of the metric (including those of the
gauge fields arising from the dimensional
reduction of the metric on squashed $S^3$) 
can be found in a similar way.
Eq.\refb{ebetarexp} gives $\beta_m=5/2$ for $d=3$. 
On the other hand since 
the five dimensional metric leads
to $SU(2)\times U(1)$ 
gauge fields and a metric on $AdS_2$ after dimensional
reduction on the squashed sphere, we see, after using the results
in \S\ref{snumzero}, that
$2\pi \int d^d y G(y) K^m(y) 
= 4 + 3=7$.
Eq.\refb{edeltasbh} now shows that the logarithmic correction
to $S_{macro}$ from the five dimensional metric is given
by $-(21/2)\ln a = -(21/4)\ln \Lambda$.

Finally we turn to the contribution due to the zero modes of
the gravitino field $\psi_\mu$. In this case eq.\refb{ebetarexp} gives 
$\beta_f=4$
for $d=3$.   On the other hand the
analysis in \S\ref{snumzero} shows that
$-2\pi\, \int d^d y \, G(y)\, 
\bar K^f(y) = 4$. Eq.\refb{edeltasbhfermi} now shows that the
net logarithmic correction to the entropy from the gravitino
zero modes is given by $3 \times 4\ln a=6\ln\Lambda$.

Combining the contributions from the vector, metric and the gravitino
zero modes we get a net logarithmic correction of
\be \label{enetlogcor}
\left[-{n_V\over 4} - {21\over 4} + 6\right] \ln\Lambda 
= -{1\over 4} (n_V-3) \ln \Lambda\, .
\ee
As already discussed, this logarithmic correction refers to the logarithm of the
index $\wt d(n,Q_1,Q_5,J)$ from the macroscopic side. 
We shall denote this index by $\wt d_{macro}(n,Q_1,Q_5,J)$.
It now follows from \refb{eshbbmpv} 
and \refb{edjdiff}
that logarithmic correction to $d_{macro}(n,Q_1,Q_5,J)$ takes the form
\be \label{enewlog}
d_{macro}(n,Q_1,Q_5,J) = e^{S_{BH}(J) -{1\over 4} (n_V-3) \ln \Lambda
+\cdots} - e^{S_{BH}(J+2) -{1\over 4} (n_V-3) \ln \Lambda
+\cdots} = e^{S_{BH}(J) -{1\over 4} (n_V-3) \ln \Lambda
+\cdots}
\ee
when the charges and the angular momentum scale as in
\refb{escalinglimit}. Here $\cdots$ denote terms of order 1.
In the term on the right hand side we have included an additional
factor of order unity, involving the ratio $J/\sqrt{Q_1Q_5\wt n}\sim 1$, 
in the $\cdots$.
\refb{enewlog}
is in perfect agreement with the microscopic result \refb{eresfin}.
It is also in agreement with \refb{enetiib} for $\alpha=0$ and
$n_V=27$, the latter being the number of vector fields
in type IIB
string theory compactified on $T^5$. By comparing \refb{enetlogcor}
and \refb{enewlog} we also see that in this limit the logarithmic
corrections to $d$ and $\wt d$ are identical.

If we take $J=0$ instead of $J\sim \Lambda^{3/2}$
then we get two extra massless gauge fields on $AdS_2$ from the
reduction of the metric on $S^3$,  giving the total 
contribution to  
$2\pi\, \int d^d y \, G(y)\, 
\bar K^m(y)$ from the metric to be $6+3=9$. 
This changes the metric
contribution of $-21/4$ in \refb{enetlogcor} to $-27/4$, and we get the
following result for the logarithmic correction to the entropy:
\be \label{ej=0macro}
-{1\over 4} (n_V+3) \ln \Lambda\, .
\ee
Furthermore this now directly computes the logarithmic correction to the
index $d(n,Q_1,Q_5,J)$.
\refb{ej=0macro} 
is in perfect agreement with the microscopic result \refb{ej=0micro}.
This is also in agreement with the microscopic result \refb{enetze}
for type IIB string theory on $T^5$ if we set $n_V=27$.

\sectiono{Microscopic analysis of the BMPV black hole entropy} 
\label{sbmpvmicro}

In this section we shall derive the
microscopic formul\ae\ for the index of a
black hole in type IIB string theory compactified
on $K3\times S^1/\ZZZ_{\bf N}$ for $\bN=1,2,3,5,7$ 
in various limits. Here  the $\ZZZ_{\bf N}$ 
symmetry acts by $2\pi/{\bf N}$
units of translation along $S^1$ and by a geometric transformation on
$K3$ that commutes with 16 supercharges. We also repeat the
analysis for type IIB string theory on $T^5$.

\subsection{Expression for the index in type IIB on
$K3\times S^1/\ZZZ_{\bf N}$} \label{eexpind}

As in \S\ref{sbmpv}
we consider
a  system of $Q_5$ D5-branes wrapped on
$K3\times S^1/\ZZZ_N$, carrying 
$Q_1$ units of D1-brane charge wrapped on $S^1$,
$-n/{\bf N}$ units of
momentum
along
$S^1$ and $SU(2)_L$ angular momentum $J_{3L}=J/2$.
We first consider the microscopic index 
\be \label{edefindalt}
\wt d_{micro}(n,Q_1,Q_5,J)\equiv
-{1\over 2!}\,
\wt{Tr}\left[ (-1)^{2 J_{3R}} \, (2 J_{3R})^2
\right]\, ,
\ee
where the trace is  taken over states 
carrying fixed $Q_1$, 
$Q_5$, $n$
and $J_{3L}=J/2$,  but
different values of $\vec J_L^2$, $J_{3R}$ and $\vec J_R^2$.
The expression for $\wt d_{micro}(n,Q_1,Q_5,J)$
can be obtained from the known expression for the elliptic genus
of the D1-D5 conformal field theory\cite{9608096}. It will however be
convenient for us to begin with the
expression
for the index of quarter BPS states in the four dimensional
theory obtained by compactifying type IIB string theory
on $K3\times 
S^1\times \wt S^1/\ZZZ_{\bf N}$\cite{9607026,0510147,0605210}, 
and then use the fact that the latter is
given by placing the five dimensional 
system we want in the background of a Kaluza-Klein
monopole associated with the $\wt S^1$ compactification\cite{0503217}. 
Thus we simply
need to remove\cite{0807.0237,0807.1314} 
from the index of the four dimensional
black hole computed in \cite{9607026,0510147,0605210}
the 
contribution of the Kaluza-Klein monopole,
and the
contribution from the supersymmetric quantum mechanics that
binds the D1-D5 system to the Kaluza-Klein monopole,
and then
multiply this by the contribution from some additional fermion
zero modes which are present in the five dimensional 
system\cite{0901.0359}.
This gives the microscopic index of such states to be (the $\bN=1$
result was written down explicitly in \cite{1009.3226})
\be \label{eintrep}
(-1)^J\wt d_{micro}(n,Q_1,Q_5,J) =
- {1\over {\bf N}} \int d\wt\rho \int d\wt\sigma \int d\wt v \,
e^{-2\pi i (\wt\rho n  + \wt\sigma Q/{\bf N} + \wt v J)} \, 
(e^{\pi i \wt v}-
e^{-\pi i \wt v})^4 {f_1({\bf N}\wt\rho)\over \wt\Phi(\wt\rho,\wt\sigma,\wt v)} \, ,
\ee
where $\wt\Phi$ is a known function of its
arguments\cite{0510147,0602254,0603066,0605210} and
can be found {\it e.g.} in eqs.(C.18) of the review 
\cite{0708.1270},
\be \label{edeff1only}
f_1(\bN\wt\rho) = \eta(\wt\rho)^{k+2} \eta({\bf N}\wt\rho)^{k+2}, \qquad
k+2 \equiv {24\over {\bf N}+1} = {1\over 2} (n_V-3)\, ,
\ee
\be \label{edefQ}
Q \equiv Q_1Q_5\, .
\ee
The contour integration over the complex variables $(\wt\rho,\wt\sigma,
\wt v)$ in \refb{eintrep}
runs along the real axes in the range $(0,1)$, $(0,{\bf N})$ and $(0,1)$
respectively at fixed values of ${\rm Im}(\wt\rho,\wt\sigma,\wt v)$.
The $(-1)^J$ factor on the left hand side of \refb{eintrep} 
can be traced
to the fact that the index is normally defined with a $(-1)^{2J_{3L}
+2 J_{3R}}$ factor inserted into the trace whereas in the definition
of $\wt d_{micro}$ we have just inserted $(-1)^{2 J_{3R}}$ into the
trace.
$-1/\wt \Phi$ is the partition function for the four dimensional
index. 
On the other hand
$1/f_1({\bf N}\wt\rho)$ is the partition function of the index associated
with the Kaluza-Klein monopole, and a factor of 
$-(e^{\pi i \wt v}-
e^{-\pi i \wt v})^{-2}$ represents the partition function associated with the
supersymmetric quantum mechanics that describes the D1-D5 center
of mass motion in the KK monopole background. Both these factors must
be removed from the four dimensional partition function 
$-1/\wt\Phi$\cite{0807.0237,0807.1314}, accounting for a
multiplicative factor of $-(e^{\pi i \wt v}-
e^{-\pi i \wt v})^2 f_1({\bf N}\wt\rho)$ in the integrand.
Another multiplicative factor of
$-(e^{\pi i \wt v}-
e^{-\pi i \wt v})^{2}$ in \refb{eintrep}
represents the index associated with the fermion
zero modes of the D1-D5 system 
carrying non-trivial $SU(2)_L$ quantum 
numbers\cite{0901.0359,1009.3226}.

Using \refb{edjdiff} for the microscopic index we can
compute the index
$d_{micro}(n,Q_1, Q_5, J)$ where we also fix
$\vec J_L^2 = {J\over 2} \left({J\over 2}+1\right)$. This gives
\ben \label{ediff}
&& d_{micro}(n,Q_1, Q_5, J)
= \wt d_{micro}(n,Q_1, Q_5, J) - \wt d_{micro}(n,Q_1, Q_5, J+2) \nn
&=& (-1)^{J+1} {1\over {\bf N}} \int d\wt\rho \int d\wt\sigma \int d\wt v \,
e^{-2\pi i (\wt\rho n  + \wt\sigma Q/{\bf N} + \wt v J)} \, 
(e^{\pi i \wt v}-
e^{-\pi i \wt v})^4  (1 - e^{-4\pi i \wt v})
{f_1({\bf N}\wt\rho)\over \wt\Phi(\wt\rho,\wt\sigma,\wt v)} \, . \nn
\een

\subsection{Evaluation of the index} \label{seval}

\refb{ediff} may be evaluated by deforming the contours of
integration of $(\wt\rho,\wt\sigma,\wt v)$
and picking up contributions from the
residues at various poles. The leading
contribution comes from the 
residue at\cite{9607026,0412287,0510147,0605210}\footnote{This 
was 
originally
derived for the limit in which $Q_1Q_5$ and $n$ scale in the same way,
and $J$ scales at either the same rate or slower than $Q_1Q_5$ and $n$.
But a careful analysis shows that this pole also gives the dominant
contribution in other scaling limits\cite{1009.3226}.}
\be\label{epolepre}
\wt\rho\wt\sigma -\wt v^2 +\wt v=0 \, ,
\ee 
where $\wt\Phi$ has a zero.
We now make a change of variables
\be\label{e6narep}
   \wt\rho={1\over {\bf N}}\, 
   {1\over 2v-\rho-\sigma}, \qquad
   \wt\sigma = {\bf N}\,
   {v^2-\rho\sigma \over 2v-\rho-\sigma}, \qquad
    \wt v =
   {v-\rho \over 2v-\rho-\sigma}\, ,
\ee
or equivalently
\be\label{e5nrep}
\rho 
   = {\wt \rho \wt\sigma - \wt v^2\over {\bf N}\wt\rho}, 
   \qquad \sigma = {\wt\rho \wt \sigma - (\wt v - 1)^2\over  
   {\bf N}\wt\rho}, \qquad
   v 
=   {\wt\rho \wt\sigma - \wt v^2 + \wt v\over {\bf N}\wt\rho}\, .
\ee
In these variables we have
\be\label{ejac}
d\wrh \wedge d\ws \wedge d\wv = -(2v -\rho-\sigma)^{-3} 
d\rho \wedge d\sigma \wedge d v\, ,
\ee
and the pole at \refb{epolepre} is situated at
\be \label{epole}
v=0\, .
\ee
Furthermore $\wt\Phi$ satisfies (see {\it e.g.} 
eq.(C.21) of the review \cite{0708.1270} where also all the other
properties of $\wt\Phi$ and $\wh\Phi$
discussed here can be found)
\be\label{enn11rep}
\wt\Phi(\wt\rho,\wt\sigma,\wt v)=-(i)^k\, C_1\, 
(2v -\rho-\sigma)^k\, \wh\Phi(\rho,\sigma,v)
\ee
where $C_1$ is a real positive
constant, $\wh\Phi(\rho,\sigma,v)$ is a new
function defined {\it e.g} in eq.(C.19) of \cite{0708.1270},
and $k$ has been defined in \refb{edeff1only}.
Using these relations in \refb{ediff} we get the leading contribution to
$d_{micro}$ to be
\bea{enn17a}
d_{micro}(n,Q_1,Q_5,J)&\simeq&  (-1)^{J+1}\,
{(i)^{-k}\over {\bf N}\, C_1}
\int_{\CC'}  d\rho \wedge   d\sigma \wedge  
dv \, (2v -\rho-
\sigma)^{-k-3} \, {1\over \wh\Phi(\rho, \sigma,  v)} \nonumber \\
&& e^{ -2\pi i(\wt\rho n  + \wt\sigma Q/{\bf N} + \wt v J)}
(e^{\pi i \wt v}-
e^{-\pi i \wt v})^4  (1 - e^{-4\pi i \wt v})
f_1(\bN\wt\rho) 
\eea
where $\CC'$ denotes a contour around $v=0$. 

Now near $v=0$ $\wh\Phi(\rho,\sigma,v)$ behaves 
as\cite{0510147,0605210,0708.1270}
\be\label{enn12rep}
\wh\Phi(\rho, \sigma,  v)=   -4\pi^2 \, v^2 \, g( \rho)\, 
g( \sigma) + \OO(  v^4)\, ,
\ee
where 
\be \label{edefgonly}
g(\rho) = \eta(\rho)^{k+2} \eta({\bf N}\rho)^{k+2}\, .
\ee
This allows us to evaluate the integration over $v$ in \refb{enn17a}
using the residue theorem. Making a further change of variables
\be\label{er3}
\rho = \tau_1+i\tau_2, \qquad \sigma = -\tau_1+i\tau_2\, ,
\ee
so that
near the pole $v=0$ we have
\ben \label{etilderel}
&& \wt\rho = {i\over 2 \bN\tau_2}\left( 1 -i {v\over \tau_2}+\OO(v^2)\right), 
\qquad \wt\sigma = i\bN {\tau_1^2 +\tau_2^2
\over 2\tau_2}\left( 1 -i {v\over \tau_2}+\OO(v^2)\right), 
\nn && 
\wt v = \left({1\over 2} - i{\tau_1\over 2\tau_2}\right)
\left( 1 -i {v\over \tau_2} - {v\over \tau_1+i\tau_2}
+\OO(v^2)\right)\, ,
\een
we can express \refb{enn17a} as
\be \label{efcont}
d_{micro}(n,Q_1, Q_5, J) \simeq
\int{d^2 \tau\over \tau_2^2} \, G(\tau_1, \tau_2)
\, ,
\ee
where
\ben \label{edefFgen}
G(\tau_1, \tau_2)
&=&  \exp\left[{\pi\over \tau_2} \left\{ {n\over {\bf N}} 
+ Q(\tau_1^2 + \tau_2^2)
- \tau_1 J \right\}\right]
\left\{ g(\tau_1+i\tau_2)
g(-\tau_1 +i\tau_2) \right\}^{-1} (2\tau_2)^{-(n_V-3)/2} 
\nonumber \\
&& \times  f_1\left({i\over 2\tau_2}\right)
\left\{2\cosh\left({\pi\tau_1\over 2\tau_2}\right)
\right\}^4 \left(1 - e^{-2\pi\tau_1/\tau_2}\right)
\nonumber \\ &&
\times \left\{ n_V-1 + {2\pi\over \tau_2}
\left( {n\over {\bf N}} + Q(\tau_1^2 + \tau_2^2)
- \tau_1 J\right) + i{1\over \tau_2}\, {f_1'(i/2\tau_2)\over
f_1(i/2\tau_2)} + 4\pi {\tau_1\over \tau_2}
\tanh{\pi\tau_1\over 2\tau_2} \right.\nn &&
\left. + 4\pi {\tau_1\over \tau_2}
{e^{-2\pi\tau_1/\tau_2}\over 1 - e^{-2\pi\tau_1/\tau_2}}\right\}  
\times \hbox{constant}\, .
\een
$\simeq$ in \refb{efcont} implies equality up to exponentially
suppressed contributions. If we had considered the index
$\wt d_{micro}(n,Q_1, Q_5, J)$ then the factor of
$\left(1 - e^{-2\pi\tau_1/\tau_2}\right)$ from the second line
and the $4\pi {\tau_1\over \tau_2}
{e^{-2\pi\tau_1/\tau_2}\over 1 - e^{-2\pi\tau_1/\tau_2}}$ factor from
the last line of \refb{edefFgen} will be absent.

We now consider the limit $n\sim Q_1\sim Q_5\sim \Lambda$,
$J^2 \sim \Lambda^3$, with
large $\Lambda$. In this case $Q\equiv Q_1Q_5\sim\Lambda^2$
and the leading contribution to \refb{efcont}
as well as systematic corrections to this formula can be found using
saddle point method. The 
saddle point values of $\tau_1$,
$\tau_2$ at the leading order, 
obtained by extremizing the exponent of the first exponential
in \refb{edefFgen} with respect to
$\tau_1$, $\tau_2$, are given by
\be \label{etau1tau2}
\tau_1 = {J\over 2Q}\sim \Lambda^{-1/2}, \quad 
\tau_2 = \sqrt{{n\over {\bf N}} - {J^2 \over 4Q}\over Q}
\sim \Lambda^{-1/2}\, .
\ee
Substituting this into the exponential of the first exponent in
\refb{edefFgen} we get the result\break 
\noindent $2\pi \sqrt{Q_1 Q_5 {n\over \bN}
-{J^2\over 4}}$ which is the leading contribution to the entropy
given in \refb{eshbbmpv}.

In order to calculate the  logarithmic corrections to
the entropy we first note that the integration over the $\tau_1$,
$\tau_2$ coordinates run along the imaginary $\tau_1$, $\tau_2$
directions\cite{0708.1270}. In these directions the first 
exponential term
in \refb{edefFgen} is sharpely peaked around the saddle point
with a width of order
\be \label{ewidth1}
\Delta \tau_1\sim \Lambda^{-5/4}, \quad \Delta\tau_2 \sim
\Lambda^{-5/4}\, ,
\ee
which can be found by studying the second derivative of the term
in the exponent with respect to $(\tau_1,\tau_2)$.
The logarithmic corrections to the log of the index
come from the following factors in \refb{efcont}, 
\refb{edefFgen} containing
powers of $\Lambda$:
\ben \label{elogsum}
(\tau_2)^{-2} &:& \Lambda \nn 
\Delta \tau_1 &:& \Lambda^{-5/4} \nn
\Delta\tau_2 &:& \Lambda^{-5/4} \nn
\{g(\tau) g(-\bar\tau)\}^{-1}
 (2\tau_2)^{-(n_V-3)/2} &:& \Lambda^{-(n_V-3)/4} \nn
\left\{ n_V-1 + {2\pi\over \tau_2}
\left( {n\over {\bf N}} + Q(\tau_1^2 + \tau_2^2)
- \tau_1 J\right) + i{1\over \tau_2}\, {f_1'(i/2\tau_2)\over
f_1(i/2\tau_2)} + 4\pi {\tau_1\over \tau_2}
\tanh{\pi\tau_1\over 2\tau_2} \right.\nn 
\left. + 4\pi {\tau_1\over \tau_2}
{e^{-2\pi\tau_1/\tau_2}\over 1 - e^{-2\pi\tau_1/\tau_2}}\right\}  &:& \Lambda^{3/2}
\, . \nn
\een
Note that in order to extract the small $\tau$ behaviour of $g(\tau)$,
given in the fourth line of \refb{elogsum},
we need to make use of the modular properties of the $\eta$-functions
appearing in \refb{edefgonly}.
Multiplying the various factors in \refb{elogsum} 
we get a net contribution of
\be\label{eresfin}
\Lambda^{-(n_V-3)/4} = \exp\left[ -{1\over 4} (n_V-3)\ln\Lambda\right]\, .
\ee
The term in the exponent on the right hand side of \refb{eresfin}
is the net logarithmic correction to $\ln d_{micro}$. 
If instead of considering
the index $d_{micro}$ we had analyzed $\wt d_{micro}$ 
we would have gotten
an identical result since the extra factors mentioned below
\refb{edefFgen}, which distinguish $d_{micro}$ from $\wt d_{micro}$,
do not contribute to the logarithmic corrections.
The result \refb{eresfin} for $d_{micro}$
is in perfect agreement with the
macroscopic result \refb{enewlog}. 

We can also consider the case when $J\sim \Lambda^{(3/2)-\alpha}$ for
some positive constant $\alpha$. In this case it follows from
\refb{etau1tau2} that 
$\tau_1\sim \Lambda^{-(1/2)-\alpha}$, $\tau_2\sim\Lambda^{-1/2}$
and hence
an additional logarithmic correction
comes from the following factor in \refb{edefFgen}
\be \label{eaddilog}
\left(1 - e^{-2\pi\tau_1/\tau_2}\right) \quad : \quad \Lambda^{-\alpha}
\ee
Thus we get the net power of $\Lambda$ in the expression for
$d(n,Q_1,Q_5,J)$ to be
\be \label{ewegeta}
 \Lambda^{-\alpha -(n_V-3)/4}
=  \exp\left[  -{1\over 4} (n_V-3 +4\alpha)\ln\Lambda\right]\, .
\ee
This agrees with the macroscopic result \refb{enewlogslow}.
If we had considered the index $\wt d_{micro}$
we would get the result \refb{eresfin} 
since the factor given in
\refb{eaddilog} is absent from the expression for $\wt d_{micro}$.

The above analysis needs some modification when $J$ vanishes
exactly since the factor $\left(1 - e^{-2\pi\tau_1/\tau_2}\right)$ vanishes
at the saddle point. In this case we proceed by expanding this in a
power series in $\tau_1$:
\be \label{epower}
\left(1 - e^{-2\pi\tau_1/\tau_2}\right) = 2\pi {\tau_1\over \tau_2}
- 2\pi^2 \left( {\tau_1\over \tau_2}\right)^2 +\cdots
\ee
Substituting this into \refb{efcont}, \refb{edefFgen} we see that the
contribution from the
term linear in $\tau_1$ vanishes by $\tau_1\to -\tau_1$ symmetry of the
rest of the integrand.\footnote{For this we need to ignore the last term inside
the curly bracket in \refb{edefFgen} which is in any case subdominant.} 
The term proportional to $(\tau_1)^2$ can be
evaluated by taking the $\tau_1$ integral to be
approximately a gaussian with
a width of order $\Delta\tau_1\sim
\Lambda^{-5/4}$ around the saddle point. Thus
$(\tau_1)^2$ factor can be replaced by a term of order $\Lambda^{-5/2}$.
On the other hand since the saddle point value of $\tau_2$ is of order
$\Lambda^{-1/2}$ which is larger than the width of the gaussian 
$\Lambda^{-5/4}$, we can replace the $(\tau_2)^2$ in the denominator by
its saddle point value of order $\Lambda^{-1}$. This gives a net factor
of $\Lambda^{-5/2}/\Lambda^{-1}\sim
\Lambda^{-3/2}$, and we get the net logarithmic correction to
$d_{micro}$ for $J=0$
to be
\be \label{ej=0micro}
\Lambda^{-3/2 -(n_V-3)/4}
=  \exp\left[  -{1\over 4} (n_V+3)\ln\Lambda\right]\, ,
\ee
in agreement with the macroscopic result \refb{ej=0macro}. 
We now see that
if we had used the index $\wt d_{micro}$ we would get the result
\refb{eresfin} which will disagree with the macroscopic result. But as
discussed in \S\ref{sensemble} in this case $d_{micro}$ is the correct index
to compare.

Before concluding this section we would like to mention that the results
for the microscopic index exist for a more general class of $\NN=4$
supersymmetric theories obtained by taking $\ZZZ_\bN$ orbifolds of
type IIB string theory on $K3\times S^1$ with non-prime $\bN$,
as well as orbifolds of type IIB string theory on $T^5$. For these models
the formula for the index takes a form similar to \refb{eintrep} and the relation
between $k$ and $n_V$ given in \refb{edeff1only} still holds although the
relation between $\bN$ and $n_V$ given in \refb{edeff1only} is lost. Also the
functions $g(\tau)$ and $f_1(\tau)$ have more complicated form, but
their large and small $\tau$ behavior are identical to what we have
discussed. Thus the results \refb{eresfin}, \refb{ewegeta} and 
\refb{ej=0micro} hold for these models as well.

\subsection{Removal of the additional hair contribution} \label{saddhair}

It has been argued in \cite{0901.0359,0907.0593} that some
hair degrees of freedom may live outside the horizon of the black
hole and hence their contribution must be
removed from the microscopic partition function before we can
compare the results to the macroscopic index associated with the
horizon degrees of freedom. The hair modes for a BMPV black hole
in type IIB string theory on $K3\times S^1$ were analyzed in
\cite{0907.0593}. A similar analysis is also possible for the orbifold models.
It was found in \cite{0907.0593} that the hair mode contribution to the
partition function includes a factor of $(e^{\pi i \wt v}-
e^{-\pi i \wt v})^{4}$ associated with the fermion zero modes 
carrying $J_{3L}$ charge and
also a $\wt\rho$ dependent factor $h(\wt\rho)$ 
associated with the modes of the
gravitino field. Thus we must multiply the integrand in \refb{ediff}
by a factor of $(e^{\pi i \wt v}-
e^{-\pi i \wt v})^{-4} (h(\wt\rho))^{-1}$. Now it follows from
\refb{etilderel}
that
at the saddle point
\be \label{esaddlept}
(e^{\pi i \wt v}-
e^{-\pi i \wt v}) = i \cosh{\pi \tau_1\over 2\tau_2}, 
\qquad \wt\rho = {i\over 2\bN \tau_2}\, .
\ee
Using \refb{etau1tau2} we see that $(e^{\pi i \wt v}-
e^{-\pi i \wt v})$ remains finite at the saddle point. On the
other hand since $\wt\rho \sim i \Lambda^{1/2}$, and since
$h(\wt\rho)$ is made of products of
$(1-e^{2\pi i \ell\wt\rho})$\cite{0907.0593},\footnote{For 
type IIB on $K3\times S^1$
we have $h(\wt\rho)=\prod_{\ell=1}^\infty (1 - e^{2\pi i\ell
\wt\rho})^{4}$.} 
it does not give any factor
involving powers of $\Lambda$. 
Thus removal of the hair contribution from the partition function
does not introduce any new
logarithmic correction to the entropy.

\subsection{BMPV black hole in type IIB on $T^5$} \label{stypeii}

We shall now briefly discuss the microscopic computation of the index
in type IIB string theory on $T^5$. In this case the known microscopic 
index 
is\cite{9903163,1009.3226}
\be \label{erelind}
\wt d_{micro}(n,Q_1,Q_5,J) = -{1\over 6!} \, 
\wt{Tr}\left[ (-1)^{2 J_{3R}} (2 J_{3R})^6
\right]\, ,
\ee
where, as before, we take the trace over states with fixed $Q_1$, $Q_5$,
$n$ and $J_{3L}=J/2$, but all $J_{3R}$, $\vec J_R^2$ and $\vec J_L^2$.
Explicit computation gives this index to be\cite{9903163}
\be \label{etx5}
\wt d_{micro}(n,Q_1,Q_5,J) \simeq (-1)^{J}\,
\int_0^1 d\tau \, \int_0^1
\, dv\, e^{-2\pi i Q_1Q_5 n \tau-2\pi i J v}
\, (e^{\pi i v} - e^{-\pi i v})^{4}\,
{\vt_1(v|\tau)^2\over \eta(\tau)^6}\, ,
\ee
up to exponentially suppressed terms. As in
\refb{ediff} we can find the index $d_{micro}$ 
for fixed $Q_1$, $Q_5$, $n$, $J_{3R}=J/2$ and
$\vec J_R^2 ={J\over 2}\left( {J\over 2}+1\right)$ by inserting in the
integrand a factor of $(1-e^{-4\pi i v})$:
\ben \label{etx5wt}
&& d_{micro}(n,Q_1,Q_5,J) = \wt d_{micro}(n,Q_1,Q_5,J)
- \wt d_{micro}(n,Q_1,Q_5,J+2)
\nn &\simeq& (-1)^{J}\,
\int_0^1 d\tau \, \int_0^1
\, dv\, e^{-2\pi i Q_1Q_5 n \tau-2\pi i J v}
\, (e^{\pi i v} - e^{-\pi i v})^{4}\,
{\vt_1(v|\tau)^2\over \eta(\tau)^6} (1 - e^{-4\pi i v})\, .
\een
We now consider the scaling limit:
\be \label{eiibsca}
Q_1,Q_5,n\sim\Lambda, \quad J\sim \Lambda^{{3\over 2}-\alpha}\, ,
\ee
and try to evaluate the integral using saddle point method.
Anticipating that at the saddle point $\tau$ is small and $v\sim 1$,
we can approximate the integral by
\ben \label{eappr}
d_{micro}(n,Q_1,Q_5,J) &\simeq& (-1)^{J}\,
\int_0^1 d\tau \, \int_0^1\, dv\,
e^{-2\pi i Q_1Q_5 n \tau-2\pi i J v}
\, (e^{\pi i v} - e^{-\pi i v})^{4}\,  e^{-2\pi i v^2/\tau}\nn && 
\qquad \qquad \qquad
e^{2\pi i v/\tau}\,
(1 - e^{-2i\pi v/\tau})^2
\, (-i\tau)^2  \, (1 - e^{-4\pi i v})\, .
\een
Extremizing the integrand with respect to $v$ and $\tau$ we find the
approximate saddle point in the range
$0\le Re(v)<1$ at\cite{1009.3226}
\be \label{eap2}
v = {1\over 2} -{J\over 2} \tau +\cdots, \qquad 
\tau = i/\sqrt{4nQ_1Q_5
-{J^2
}}+\cdots \sim \Lambda^{-3/2} \, ,
\ee
where $\cdots$ denote subleading terms. 
The value of the integrand
at this saddle point is
\be \label{eap3}
\exp[\pi \sqrt{4nQ_1Q_5 - J^2} +\cdots]\, .
\ee
This gives the leading contribution to $d_{micro}(n,Q_1,Q_5,J)$.
To examine the logarithmic corrections we determine the various
powers of $\Lambda$ coming from different terms in \refb{eappr}:
\ben \label{epowera}
\Delta v &:&  \Lambda^{-3/4} \nn
\Delta\tau &:& \Lambda^{-9/4} \nn
(-i\, \tau)^2 &:& \Lambda^{-3} \nn
(1 - e^{-4\pi i v}) &:& \Lambda^{-\alpha} 
\een
where $\Delta v$ and $\Delta \tau$ denotes the range of $v$ and $\tau$
integration beyond which the integrand falls off sharply. Taking the
product of these factors we get the net power of $\Lambda$ in the
expression for the index:
\be \label{enetiib}
\Lambda^{-6-\alpha} = \exp[-(6+\alpha)\ln \Lambda]\, .
\ee
For $\alpha=0$
this agrees with the macroscopic result \refb{enewlog} for 
$n_V=27$. On the other hand for $\alpha>0$,
\refb{enetiib} agrees with the macroscopic result \refb{enewlogslow}.
For computing the index $\wt d_{micro}$
we need to 
drop the $(1 - e^{-4\pi i v})$ from the integrand in
\refb{eappr} with no other change.
According to \refb{epowera}, we now get the result
\be \label{enetiibal0}
\Lambda^{-6} = \exp[-6\ln \Lambda]\, .
\ee

If we instead set $J=0$ then the relevant index for comparison
with the macroscopic result  $d_{micro}$. In this case
special care is needed to deal with the
factor of $(1 - e^{-4\pi i v})$. Since this vanishes at the saddle point,
we expand it to second order in fluctuations about the saddle point
and replace it by a term of order $\Delta v^2 \sim \Lambda^{-3/2}$.
This gives a net contribution of
\be \label{enetze}
\Lambda^{-6-3/2} = \exp[-{15\over 2} \ln \Lambda]\, ,
\ee
to $d_{micro}$.
Again this agrees with the macroscopic result \refb{ej=0macro}
for $n_V=27$. However if we had considered the index $\wt d_{micro}$
we would have gotten the result \refb{enetiibal0} which would
disagree with the macroscopic result. This again illustrates the necessity
for making the correct choice of ensembles while comparing the
microscpic and the macroscopic results.

\subsection{Comparison with the Cardy limit} \label{scardy}

Since the success of the black hole microstate counting is often
associated with the Cardy formula -- applicable for $n\to\infty$
limit at fixed $Q_1$, $Q_5$ -- we shall analyze in this subsection the
logarithmic correction to the entropy in the Cardy limit. For this
we consider the scaling
\be \label{ecardyscaling}
n\sim \Lambda^3, \qquad J\sim \Lambda^{3/2-\alpha},
\ee
with $Q_1$, $Q_5$ fixed. Eq.\refb{etau1tau2} shows that in
this limit
\be \label{ecas2}
\tau_2\sim \Lambda^{3/2}, \qquad \tau_1 \sim \Lambda^{3/2-\alpha}\, .
\ee
Thus the logarithmic correction to $\ln d_{micro}$ comes from
the terms:
\ben \label{ecard1}
(\tau_2)^{-2} &:& \Lambda^{-3} \nn 
\Delta \tau_1 &:& \Lambda^{3/4} \nn
\Delta\tau_2 &:& \Lambda^{3/4} \nn
g(\tau) g(-\bar\tau) (2\tau_2)^{-(n_V-3)/2} &:& \Lambda^{-3(n_V-3)/4} \nn
f_1(i/2\tau_2) &:& \Lambda^{3(n_V-3)/4} \nn
\left\{ n_V-1 + {2\pi\over \tau_2}
\left( {n\over {\bf N}} + Q(\tau_1^2 + \tau_2^2)
- \tau_1 J\right) + i{1\over \tau_2}\, {f_1'(i/2\tau_2)\over
f_1(i/2\tau_2)} + 4\pi {\tau_1\over \tau_2}
\tanh{\pi\tau_1\over 2\tau_2} \right.\nn 
\left. + 4\pi {\tau_1\over \tau_2}
{e^{-2\pi\tau_1/\tau_2}\over 1 - e^{-2\pi\tau_1/\tau_2}}\right\}  &:& \Lambda^{3/2}\nn
(1 - e^{-2\pi \tau_1/\tau_2}) &:& \Lambda^{-\alpha}
\nn
\een
Note that we now need to use the modular property of the 
$\eta$-functions appearing in \refb{edeff1only} 
to find the behavior of $f_1$ for small value of its argument.
Taking the product of all the factors we get\footnote{If we had
been considering the elliptic genus of symmetric products of K3
-- which is a weak Jacobi form of weight zero -- then for $\alpha=0$
we should have gotten a power correction of order $n^{-1}\sim
\Lambda^{-3}$\cite{0005003}. However the contribution from the
center of mass oscillation modes turns the elliptic genus into a 
(meromorphic) weak Jacobi form of weight two, and kills the
logarithmic term in the asymptotic expansion of the Fourier
coefficients.
}
\be \label{ecard2}
\Lambda^{-\alpha} =\exp[-\alpha\ln\Lambda]\, ,
\ee
which is quite different from \refb{ewegeta}. In particular there is
no dependence on $n_V$. This shows that the Cardy limit is not
always reliable for extracting the black hole entropy in the limit
where the supergravity approximation is reliable. Similar observations
have been made earlier in \cite{0807.0237,1009.3226}, and it has in fact
been argued in \cite{1009.3226} that the limit $Q_1Q_5>> n$ is
related to the Cardy limit of a different CFT that appears in the
dual type IIA description.

We can also consider a similar limit for the formula
\refb{etx5wt} for type IIB string theory on $T^5$. 
{}From \refb{etx5wt} we see that the index depends only on the
combination $Q_1Q_5n$ and not individually on $Q_1$, $Q_5$
and $n$. Thus in this case 
scaling $n$ by $\Lambda^3$ keeping $Q_1$ and $Q_5$
fixed is equivalent to 
scaling $Q_1$, $Q_5$ and $n$ by 
$\Lambda$, and we get a result identical to \refb{enetiib}:
\be \label{eiibcardy}
\exp[-(6+\alpha)\ln \Lambda]\, .
\ee

\sectiono{Slowly rotating black holes} \label{sslowrot}

In the macroscopic analysis of
\S\ref{sbmpv} we have considered two kinds of cases -- ({\it i})
when the angular
momentum is large so that its contribution to the classical
Bekenstein-Hawking entropy scales in the
same way as that from the charges, and ({\it ii}) when the
angular momentum vanishes. In this section we shall consider some
cases when the angular momentum is non-zero but parametrically
small so that it gives negligible contribution to the classical
Bekenstein-Hawking entropy. This analyis
will however require us to make
some assumptions about the spectrum of the kinetic operator in
the near horizon geometry of a slowly rotating black hole. 
For this reason the results of this section should
be regarded as somewhat tentative.

\subsection{Slowly rotating BMPV black holes} \label{sslownew}

The microscopic analysis for BMPV black holes  yields the
result for the index in the scaling limit
$Q_1,Q_5,n\sim\Lambda$, $J\sim \Lambda^{{3\over 2}
-\alpha}$ for some positive number $\alpha$ -- as given in eqs.\refb{ewegeta}
and \refb{enetiib}. 
Since $J/\Lambda^{3/2}$ is the parameter that controls the deviation of the
black hole solution from the rotationally invariant configuration, our first
guess will be that the partition function on the macroscopic side can be
computed 
using perturbation expansion in
$J/\Lambda^{3/2}$ around the rotationally invariant configuration $J=0$.
Furthermore,
one would naively expect that the
corrections to the non-zero eigenvalues would
have negligible effect for small $J/\Lambda^{3\over 2}$ 
and we only need to take into account the shift in the
zero eigenvalues using perturbation theory. This would produce small
but non-zero eigenvalues, and, as a result,
compared to the case of $J\sim \Lambda^{3/2}$, the one loop determinant
will get some extra power of $J/\Lambda^{3\over 2}$ from the
small eigenvalues. 
However this naive expectation is not quite 
correct due to the following reason. 
We recall that the zero modes which are lifted by switching
on $J$ are the ones associated with the $W^\pm_L$ gauge fields on
$AdS_2$ arising from $SU(2)_L$.
Now
a non-zero $J$ corresponds
to switching on a constant $U(1)_L$ 
electric field along $AdS_2$.  Since the
$W^{\pm}_L$ fields are charged under the $U(1)_L$, any quantum of
$W^\pm_L$ will be subject to a constant force in the presence of a
$U(1)_L$ electric field. Hence we cannot use
perturbation theory to study the effect of the switching on a
$U(1)_L$ electric field on the
zero modes of $W^\pm_L$ -- we expect these zero modes to be lifted
altogether in the presence of non-zero $J$.\footnote{This is corroborated by
the fact that if in \refb{e24p} we replace $d\Phi^{(\ell)}$ by 
$D\Phi^{(\ell)}$ where $D$ denotes the gauge covariant
derivative for the $W^\pm_L$ fields in the background of
$U(1)_L$ electric field, then the modes
cease to be square integrable.}
While we do not have a concrete analysis of the eigenvalue equations
in the new background, it is natural
to assume that all the modes, except the
exact zero modes which are in any case neutral under $U(1)_L$, 
will have eigenvalues of order $a^{-2}$. In other words
the effect of switching on even a small $J$ takes us to back to the case
of generic $J/\Lambda^{3/2}\sim 1$, \i.e.\ gives 
a logarithmic correction to $ \ln \wt d_{macro}(n,Q_1, Q_5, J)$ given in
\refb{enetlogcor}. Using the fact that $S_{BH}$ given in
\refb{eshbbmpv} satisfies
\be \label{esbhnewform}
S_{BH}(J) - S_{BH}(J+2) \simeq {\pi\over 2} {J\over \sqrt{Q_1Q_5\wt n}}
\sim \Lambda^{-\alpha}
\ee
for $J\sim \Lambda^{{3\over 2}-\alpha}$, we see from \refb{enewlog}
that the logarithmic
correction to $d_{macro}(n,Q_1,Q_5,J)$ now takes the form:
\be \label{enewlogslow}
d_{macro}(n,Q_1,Q_5,J) = e^{S_{BH}(J) -{1\over 4} (n_V-3) \ln \Lambda
+\cdots} - e^{S_{BH}(J+2) -{1\over 4} (n_V-3) \ln \Lambda
+\cdots} = e^{S_{BH}(J) -{1\over 4} (n_V-3+4\alpha) \ln \Lambda
+\cdots}\, .
\ee
This agrees with the microscopic result \refb{ewegeta} and also with
\refb{enetiib} for $n_V=27$.

\subsection{Slowly rotating extremal Kerr-Newmann black hole in
four dimensions} \label{slow}

In this subsection we shall briefly discuss the case of 
slowly rotating extremal
Kerr-Newmann black hole in four dimensional Einstein-Maxwell
theory. For zero angular momentum, \i.e.\ for Reissner-Nordstrom 
black hole, the result for the logarithmic correction was
calculated in \cite{1108.3842} with the result:
\be \label{eresrn}
-{1\over 180} (964 + n_S + 62 n_V + 11 n_F) \, \ln {A_H}
\ee
if the theory contains, besides the metric and the Maxwell field under
which the black hole is charged, $n_S$ minimally coupled massless scalar
fields, $n_V$ minimally coupled {\it additional} massless vector fields and
$n_F$ minimally coupled massless Dirac fermions. 
Here $A_H\sim a^2$ is the area of the event horizon and we have
set the Newton's constant $G_N$ to unity.

Let us now consider the effect of deforming the solution
so that the black hole carries not only electric charge $q$ but also angular
momentum $J$ while remaining extremal. The near horizon geometry still
contains an $AdS_2$ factor\cite{9905099,0606244},
but the $SO(3)$ isometry of the near horizon geometry is now
broken to $U(1)$ and as a result from the dimensional reduction of the
metric we get only one massless gauge field on $AdS_2$ instead of
three. The departure of the
metric from that of the Reissner-Nordstrom black hole is given by the
rotation parameter
\be \label{erotpar}
\gamma \equiv {J\over q^2}\, .
\ee
We shall work with $\gamma$ small (but still $J$ large) 
so that the geometry is almost that
of the extremal Reissner-Nordstrom black hole. 
Following the discussion in 
\S\ref{sslownew} we shall assume that the effect of
switching on even a small $\gamma$ will be to get rid of the zero modes
of two of the three gauge fields coming from the unbroken $SO(3)$ isometry of
the non-rotating black holes and make all the eigenvalues, except those
associated with the exact zero modes, of order $a^{-2}$. 
In the language of this paper, this translates to the fact
that $2\pi \int d^2 y G(y) \bar K^m(y)$, instead of taking the
value $3+3=6$ will now take value $1+3=4$. Since here $\beta_m=2$,
this means according to \refb{edeltasbh} that we lose an additive
contribution of $-2\ln a = -\ln A_H$ from the entropy. Adding $\ln A_H$ to
\refb{eresrn} we get
\be \label{eresrnmod}
\wt d_{macro}(q,J) =
\exp\left[S_{BH}(q,J) -{1\over 180} (784 + n_S + 62 n_V + 11 n_F) \, \ln {A_H}
+\cdots \right]\, .
\ee
Note that we have used the fact that for non-zero angular momentum the
macroscopic computation yields the degeneracy $\wt d_{macro}(q,J)$ 
in the ensemble of fixed $J_3=J$, instead of $d_{macro}(q,J)$ which will be the
degeneracy in the ensemble of fixed $J_3=J$ and $\vec J^2=J(J+1)$.
However we can use \refb{eresrnmod} and the result for the classical
Bekenstein-Hawking entropy:
\be \label{eclassbh}
S_{BH}(q,J) = 2\pi\, \sqrt{q^4 + J^2} =
2\pi \, q^2 \left( 1 + {1\over 2} \gamma^2 + \OO(\gamma^4)\right)\, ,
\ee
to compute $d_{macro}(q,J)$:
\ben \label{edqjf}
d_{macro}(q,J) &=& \wt d_{macro}(q,J) - \wt d_{macro}(q, J+1) \nn
&\simeq& 
\exp\left[S_{BH}(q,J) -{1\over 180} (784 + n_S + 62 n_V + 11 n_F) \, \ln {A_H}
+\ln \gamma \right]\, .\nn
\een

\bigskip

{\bf Acknowledgement:} I wish to thank Monica Guica and Rajesh Gupta
for useful
communications.
This work was
supported in part by the J. C. Bose fellowship of 
the Department of Science and Technology, India and the 
project 11-R\&D-HRI-5.02-0304.

\appendix

\sectiono{Counting of zero modes in $AdS_2$} \label{sbasis}

In this appendix we shall review the results on the zero modes of
various fields in $AdS_2$ and count their numbers. 
For definiteness we shall take the $AdS_2$ metric to be
\be \label{eadsmet}
(g_{AdS_2})_{\mu\nu} dx^\mu d x^\nu =
a^2 (d\eta^2 +\sinh^2\eta d\theta^2)\, ,
\ee
although the result for the number of zero modes will be independent
of what we take to be the $AdS_2$ size $a$.

First consider
the case of a $U(1)$ gauge field $A_\mu$. The normalized
basis of zero modes of such
a field on $AdS_2$ is given
by\cite{camhig1}
\be \label{e24p}
A = d\Phi^{(\ell)}, \qquad \Phi^{(\ell)} = {1\over \sqrt{2\pi |\ell|}}\,
\left[ {\sinh\eta \over 1+\cosh\eta}\right]^{|\ell|} e^{i\ell\theta},
\quad \ell = \pm 1, \pm 2, \pm 3, \cdots\, ,
\ee
satisfying
\be \label{enormgauge}
\int_{AdS_2} d^2 x \, \sqrt{g_{AdS_2}} \, g_{AdS_2}^{mn} \p_m \Phi^{(\ell)*}
\p_n \Phi^{(\ell')} = \delta_{\ell \ell'}\, .
\ee
The basis states
\refb{e24p} also satisfy
\be \label{esumvector}
\sum_\ell g_{AdS_2}^{mn} \p_m \Phi^{(\ell)*}(x) \p_n \Phi^{(\ell)}(x)
= {1\over 2\pi a^2}\, .
\ee
This can be derived using the fact that due to homogeneity of
$AdS_2$ this sum is independent of $x$ and hence can be evaluated
at $\eta=0$, and that at $\eta=0$
only the $\ell=\pm 1$ terms contribute to the sum. Thus the 
total number
of such zero modes of a vector field on $AdS_2$ is given by
\be \label{etotalvz}
N_1 = \int_{AdS_2}\,  d^2 x \, \sqrt{g_{AdS_2}} \, 
\sum_\ell g_{AdS_2}^{mn} \p_m \Phi^{(\ell)*}(x) \p_n \Phi^{(\ell)}(x)
= {1\over 2\pi} \int_0^{\eta_0} \sinh\eta\, d\eta \, \int d\theta
= \cosh\eta_0 - 1 \, .
\ee

A similar analysis can be done for a symmetric rank two
tensor representing the graviton fluctuation on $AdS_2$. 
The normalized basis of
zero mode deformations is given by\cite{camhig1}
\ben \label{ehmn}
h_{mn}&=& w^{(\ell)}_{mn}, \nn
w^{(\ell)}_{mn} dx^m dx^n &=&
{a\over  \sqrt{\pi}} \, \left[ {  |\ell| 
(\ell^2-1)\over
2} \right]^{1/2} \, {(\sinh\eta)^{|\ell|-2} \over (1 +\cosh\eta)^{|\ell|}}
\, e^{i\ell\theta}\, (d\eta^2 + 2 \, i\, \sinh\eta\, d\eta d\theta
- \sinh^2\eta \, d\theta^2) \nn
&& \quad \ell\in \ZZZ, \quad |\ell|\ge 2\, .
\een
Locally these can be regarded as deformations generated by
a diffeomorphism on $AdS_2$, 
but these diffeomorphisms 
themselves are not square integrable. 
The basis states
\refb{ehmn} satisfy
\be \label{esummetric}
\sum_\ell g_{AdS_2}^{mn} g_{AdS_2}^{pq} 
w^{(\ell)*}_{mp}(x) w^{(\ell)}_{nq}(x)
= {3\over 2\pi a^2}\, .
\ee
We have derived this using the fact that due to homogeneity of
$AdS_2$ this sum is independent of $x$, and that at $\eta=0$
only the $\ell=\pm 2$ terms contribute to the sum.
Thus as in \refb{etotalvz}
the total number of such discrete modes is given by
\be \label{ediscretemetric}
N_2 = {3 \cosh\eta_0} - 3 \, .
\ee

Finally we turn to the zero modes of the gravitino fields.
We use the following conventions for the zweibeins  and the gamma
matrices
\be \label{evier1}
e^0= a\, \sinh\eta \, d\theta, \quad e^1 = a\, d\eta \, ,
\ee
\be \label{egam1}
\gamma^0 = -\tau_2, \quad \gamma^1 = \tau_1 ,
\ee
where  $\tau_i$ are two dimensional Pauli matrices and $a$ is the
$AdS_2$ size parameter.
In this convention 
the Dirac operator on
$AdS_2$ can be written as
\be \label{ed1a}
\not \hskip -4pt D_{AdS_2} =
a^{-1}\left[ -\tau^2\, {1\over \sinh\eta} \p_\theta
+ \tau^1 \, \p_\eta +{1\over 2}\, \tau^1\, \coth\eta\right]\, .
\ee
The eigenstates of $\not\hskip -4pt D_{AdS_2}$
are given by\cite{9505009}
\ben \label{ed2a}
\chi_{k}^{\pm}(\lambda) &=& {1\over \sqrt{4\pi a^2}}\,
\left|{\Gamma\left( {1} + k + i\lambda\right)
\over \Gamma(k+1) \Gamma\left({1\over 2}+i\lambda\right)}\right|\,
e^{ i\left(k+{1\over 2}\right)\theta}  \nn
&& \qquad 
\pmatrix{ i \, {\lambda\over k+1}\, 
\cosh^{k}{\eta\over 2}\sinh^{k+1} {\eta\over 2}
F\left(k+1+i\lambda, k+1-i\lambda; k+2;-\sinh^2{\eta\over 2}\right)
\cr
\pm \cosh^{k+1}{\eta\over 2}\sinh^k {\eta\over 2}
F\left(k+1+i\lambda, k+1-i\lambda; k+1;-\sinh^2{\eta\over 2}\right)}, \nn \cr \cr
\eta_{k}^{\pm}(\lambda) &=& {1\over \sqrt{4\pi a^2}}\,
\left|{\Gamma\left( {1} + k + i\lambda\right)
\over \Gamma(k+1) \Gamma\left({1\over 2}+i\lambda\right)}\right|\,
e^{ -i\left(k+{1\over 2}\right)\theta}\nn
&&  \qquad 
\pmatrix{ \cosh^{k+1}{\eta\over 2}\sinh^k {\eta\over 2}
F\left(k+1+i\lambda, k+1-i\lambda; k+1;-\sinh^2{\eta\over 2}\right)
\cr
\pm i \, {\lambda\over k+1}\, 
\cosh^{k}{\eta\over 2}\sinh^{k+1} {\eta\over 2}
F\left(k+1+i\lambda, k+1-i\lambda; k+2;-\sinh^2{\eta\over 2}\right)
}, 
\nn \cr && 
\qquad  k\in \ZZZ, \quad 0\le k<\infty, \quad 0<\lambda<\infty\, ,
\een
satisfying
\be \label{eadsev}
\not \hskip -4pt D_{AdS_2} \chi_{k}^{\pm}(\lambda)=\pm i\, a^{-1}\,
\lambda\,  \chi_{k}^{\pm}(\lambda)
\, , \qquad
\not \hskip -4pt D_{AdS_2} \eta_{k}^{\pm}(\lambda)
=\pm i\, a^{-1}\,
\lambda\,  \eta_{k}^{\pm}(\lambda)
\, .
\ee

The zero modes of the gravitino fields $\psi_m$ can be
expressed in terms of the spinors \refb{ed2a} via the relations
\be \label{eadd1}
\xi_m^{(k)\pm}\equiv \NN^\pm_{k}\, 
\left(D_m\pm {1\over 2a} 
\gamma_m\right)\chi^\pm_k(i),
\qquad \wh\xi_m^{(k)\pm}\equiv \wh\NN^\pm_{k}\, 
\left(D_m\pm {1\over 2a}
\gamma_m\right)
\eta^\pm_k(i), \quad k=1,\cdots \infty \, ,
\ee
where $\NN^\pm_{k}$ and $\wh\NN^\pm_{k}$ 
are appropriate normalization constants such that
\be \label{enormgt}
a^2 \int \sinh\eta\, d\eta\, d\theta \, g_{AdS_2}^{mn}\, 
\xi_m^{(k)+\dagger}(\eta,\theta) 
\xi_n^{(k')+}(\eta,\theta) =\delta_{kk'}
\ee
etc.
Although $\chi^\pm_k(i)$ and $\eta^\pm_k(i)$ are
not square integrable,
the modes described in \refb{eadd1} are square
integrable and hence they must be included among the eigenstates
of the Rarita-Schwinger operator.
These modes can be shown to satisfy the chirality projection
condition
\be \label{eimp1}
\tau_3 \, \xi_m^{(k)\pm}
= -\xi_m^{(k)\pm}, \qquad 
\tau_3  \, \wh\xi_m^{(k)\pm}
=  \wh\xi_m^{(k)\pm}\, .
\ee
Furthermore with the help of \refb{eadsev} and that 
$\chi_k^- =\tau_3\chi_k^+$, $\eta_k^-=\tau_3\eta_k^+$ 
one can show that
$\xi_m^{(k)\pm}$ are proportional to each other and
$\wh\xi_m^{(k)\pm}$ are proportional to each other.

Now suppose we have a set of gravitino zero modes given by
$\xi_m^{(k)+}$ with $k$ ranging over all positive integers. Then 
the total number of zero modes may be expressed as
\be \label{efzerot}
\sum_{k=1}^\infty 1 =
a^2 \int \sinh\eta\, d\eta\, d\theta \,  \sum_{k=1}^\infty 
g^{mn} \,
\xi_m^{(k)+\dagger}(\eta,\theta) \xi_n^{(k)+}(\eta,\theta)\, .
\ee
We now use the fact that after taking the sum over $k$
the integrand must become independent of $(\eta,\theta)$ and
hence we can evaluate it at $\eta=0$. In this case the only
contribution comes from the $k=1$ term.
Substituting $k=1$, $\lambda=i$ in \refb{ed2a} and choosing the
normalization constant $\NN^\pm_{1}$ so that
$\xi_m^{(1)+}$ defined in \refb{eadd1} is normalized, 
we get
\be \label{eexpgr1}
\xi_\theta^{(1)+} = {1\over \sqrt\pi} \, e^{{3}i\theta/2}
\pmatrix{0\cr \sinh{\eta\over 2}/
\cosh^2{\eta\over 2}}, \quad \xi_\eta^{(1)+} =
-{i\over \sinh\eta} \, \xi_\theta^{(1)+}\, .
\ee
This gives
\be \label{egives}
\sum_{k=1}^\infty 
g^{mn} \,
\xi_m^{(k)+\dagger}(\eta,\theta) \xi_n^{(k)+}(\eta,\theta)
= g^{mn} \,
\xi_m^{(1)+\dagger}(\eta,\theta) \xi_n^{(1)+}(\eta,\theta)|_{\eta=0}
= {1\over 2\pi a^2}\, .
\ee
Substituting this into \refb{efzerot} we get the total number of
zero modes to be $(\cosh\eta_0-1)$. 
A similar contribution is found from the 
gravitino zero modes given by
$\wh\xi_m^{(k)+}$.

We should however remember that the spinors $\xi_m^{(k)+}$
and $\wh\xi_m^{(k)+}$ are tensored with spinors associated with
the tangent space spinors of other directions transverse to
$AdS_2$ and hence each zero mode associated with
$\xi_m^{(k)+}$
and $\wh\xi_m^{(k)+}$ may actually represent multiple zero modes.
In order to determine this multiplicity we shall use the fact that
the gravitino zero modes are associated with the deformations
generated by the fermionic generators
of the $\NN=4$ superconformal
algebra labelled as $G^{\alpha\beta}_n$ where $\alpha$ and $\beta$
each takes value $\pm 1$ and 
$n\in \ZZZ+{1\over 2}$\cite{0903.1477,0905.2686}. 
$G^{\alpha\beta}_{\pm 1/2}$ are the global 
symmetry generators which, together with the
bosonic generators, form the $SU(1,1|2)$ 
supersymmetry of the near horizon 
geometry\cite{0608021,0905.2686,1012.0265}. In the present
context $G^{\alpha\beta}_n$ for $n\ge {3\over 2}$ can be identified with the
zero modes $\xi_m^{(k)+}$ with $n=k+{1\over 2}$ and
$G^{\alpha\beta}_n$ for $n\le -{3\over 2}$ can be identified with the
zero modes $\wh\xi_m^{(k)+}$ with $|n|=k+{1\over 2}$.
Since for each $n$ there are four generators labelled by the
pair $(\alpha,\beta)$ we see that 
the zero modes associated with $\xi_m^{(k)+}$
and $\wh\xi_m^{(k)+}$ must each have multiplicity 4. This gives
the total number of fermion zero modes to be
\be \label{etotfin}
N_{3/2} = 8(\cosh\eta_0 - 1)\, .
\ee
Explicit construction of the four zero modes for each $\xi^{(k)+}_m$
(and $\wh \xi^{(k)+}_m$) can be found in \cite{1106.0080,1108.3842})
for BPS black holes in four dimensional $\NN=2$ and $\NN=4$
supersymmetric theories.


\small

\baselineskip 10pt

\begin{thebibliography}{99}


\bibitem{9601029}
  A.~Strominger and C.~Vafa,
  ``Microscopic Origin of the Bekenstein-Hawking Entropy,''
  Phys.\ Lett.\ B {\bf 379}, 99 (1996)
  [arXiv:hep-th/9601029].

\bibitem{9602065}
  J.~C.~Breckenridge, R.~C.~Myers, A.~W.~Peet and C.~Vafa,
  ``D-branes and spinning black holes,''
  Phys.\ Lett.\  B {\bf 391}, 93 (1997)
  [arXiv:hep-th/9602065].

\bibitem{9307038}
  R.~M.~Wald,
  ``Black hole entropy in the Noether charge,''
  Phys.\ Rev.\ D {\bf 48}, 3427 (1993)
  [arXiv:gr-qc/9307038].

\bibitem{9312023}
  T.~Jacobson, G.~Kang and R.~C.~Myers,
  ``On Black Hole Entropy,''
  Phys.\ Rev.\ D {\bf 49}, 6587 (1994)
  [arXiv:gr-qc/9312023].

\bibitem{9403028}
  V.~Iyer and R.~M.~Wald,
  ``Some properties of Noether 
  charge and a proposal for dynamical black hole
  entropy,''
  Phys.\ Rev.\ D {\bf 50}, 846 (1994)
  [arXiv:gr-qc/9403028].

\bibitem{9502009}
  T.~Jacobson, G.~Kang and R.~C.~Myers,
  ``Black hole entropy in higher curvature gravity,''
  arXiv:gr-qc/9502009.

\bibitem{9812082}
G.~Lopes Cardoso, B.~de Wit and T.~Mohaupt,
``Corrections to macroscopic supersymmetric black-hole entropy,''
Phys.\ Lett.\ B {\bf 451}, 309 (1999)
[arXiv:hep-th/9812082].

\bibitem{9904005}
G.~Lopes Cardoso, B.~de Wit and T.~Mohaupt,
``Deviations from the area law for supersymmetric black holes,''
Fortsch.\ Phys.\  {\bf 48}, 49 (2000)
[arXiv:hep-th/9904005].

\bibitem{0007195}
T.~Mohaupt,
``Black hole entropy, special geometry and strings,''
Fortsch.\ Phys.\  {\bf 49}, 3 (2001)
[arXiv:hep-th/0007195].

\bibitem{9607026}
R.~Dijkgraaf, E.~P.~Verlinde and H.~L.~Verlinde,
``Counting dyons in N = 4 string theory,''
Nucl.\ Phys.\ B {\bf 484}, 543 (1997)
[arXiv:hep-th/9607026].

\bibitem{0412287}
  G.~Lopes Cardoso, B.~de Wit, J.~Kappeli and T.~Mohaupt,
  ``Asymptotic degeneracy of dyonic N = 4 string states and black hole
  entropy,''
  JHEP {\bf 0412}, 075 (2004)
  [arXiv:hep-th/0412287].

\bibitem{0505094}
D.~Shih, A.~Strominger and X.~Yin,
``Recounting dyons in N = 4 string theory,''
arXiv:hep-th/0505094.

\bibitem{0506249}
D.~Gaiotto,
``Re-recounting dyons in N = 4 string theory,''
arXiv:hep-th/0506249.

\bibitem{0508174}
  D.~Shih and X.~Yin,
  ``Exact black hole degeneracies and the topological string,''
  JHEP {\bf 0604}, 034 (2006)
  [arXiv:hep-th/0508174].

\bibitem{0510147}
  D.~P.~Jatkar and A.~Sen,
  ``Dyon spectrum in CHL models,''
  JHEP {\bf 0604}, 018 (2006)
  [arXiv:hep-th/0510147].

\bibitem{0602254}
  J.~R.~David, D.~P.~Jatkar and A.~Sen,
  ``Product representation of dyon partition function in CHL models,''
  JHEP {\bf 0606}, 064 (2006)
  [arXiv:hep-th/0602254].

\bibitem{0603066}
  A.~Dabholkar and S.~Nampuri,
  ``Spectrum of dyons and black holes in
  CHL orbifolds using Borcherds lift,''
  arXiv:hep-th/0603066.

\bibitem{0605210}
  J.~R.~David and A.~Sen,
  ``CHL dyons and statistical entropy function from D1-D5 system,''
  arXiv:hep-th/0605210.

\bibitem{0607155}
  J.~R.~David, D.~P.~Jatkar and A.~Sen,
  ``Dyon spectrum in N = 4 supersymmetric type II string theories,''
  arXiv:hep-th/0607155.

\bibitem{0609109}
  J.~R.~David, D.~P.~Jatkar and A.~Sen,
  ``Dyon spectrum in generic N = 4 supersymmetric Z(N) orbifolds,''
  arXiv:hep-th/0609109.

\bibitem{9712251}
  A.~Strominger,
  ``Black hole entropy from near-horizon microstates,''
  JHEP {\bf 9802}, 009 (1998)
  [arXiv:hep-th/9712251].

\bibitem{0809.4266}
  M.~Guica, T.~Hartman, W.~Song, A.~Strominger,
  ``The Kerr/CFT Correspondence,''
  Phys.\ Rev.\  {\bf D80}, 124008 (2009).
  [arXiv:0809.4266 [hep-th]].

\bibitem{0809.3304}
  A.~Sen,
  ``Quantum Entropy Function from AdS(2)/CFT(1) Correspondence,''
  Int.\ J.\ Mod.\ Phys.\  A {\bf 24}, 4225 (2009)
  [arXiv:0809.3304 [hep-th]].

\bibitem{1008.3801}
  I.~Mandal, A.~Sen,
  ``Black Hole Microstate Counting and its Macroscopic Counterpart,''
  Class.\ Quant.\ Grav.\  {\bf 27}, 214003 (2010).
  [arXiv:1008.3801 [hep-th]].

\bibitem{1005.3044}
  S.~Banerjee, R.~K.~Gupta, A.~Sen,
  ``Logarithmic Corrections to Extremal Black Hole Entropy from Quantum Entropy Function,''  
  JHEP {\bf 1103}, 147 (2011) [arXiv:1005.3044 [hep-th]].

\bibitem{1106.0080}
  S.~Banerjee, R.~K.~Gupta, I.~Mandal, A.~Sen,
  ``Logarithmic Corrections to N=4 and N=8 Black Hole Entropy: A One Loop Test of Quantum Gravity,''
  [arXiv:1106.0080 [hep-th]].
  
\bibitem{1108.3842}
  A.~Sen,
  ``Logarithmic Corrections to N=2 Black Hole Entropy: 
  An Infrared Window into the Microstates,''
  [arXiv:1108.3842 [hep-th]].

\bibitem{1109.0444}
S.~Ferrara and A.~Marrani,  
``Generalized Mirror Symmetry and Quantum Black Hole Entropy'',
[arXiv:1109.0444 [hep-th]].

\bibitem{0506151}
  D.~Shih, A.~Strominger and X.~Yin,
  ``Counting dyons in N = 8 string theory,''
  JHEP {\bf 0606}, 037 (2006)
  [arXiv:hep-th/0506151].

\bibitem{0903.1477}
  A.~Sen,
  ``Arithmetic of Quantum Entropy Function,''
  JHEP {\bf 0908}, 068 (2009)
  [arXiv:0903.1477 [hep-th]].

\bibitem{1009.3226}
  A.~Dabholkar, J.~Gomes, S.~Murthy, A.~Sen,
  ``Supersymmetric Index from Black Hole Entropy,''
  JHEP {\bf 1104}, 034 (2011).
  [arXiv:1009.3226 [hep-th]].

\bibitem{9608096}
  R.~Dijkgraaf, G.~W.~Moore, E.~P.~Verlinde and H.~L.~Verlinde,
  ``Elliptic genera of symmetric products and second quantized strings,''
  Commun.\ Math.\ Phys.\  {\bf 185}, 197 (1997)
  [arXiv:hep-th/9608096].

\bibitem{9505054}
S.~Chaudhuri, G.~Hockney and J.~D.~Lykken,
``Maximally supersymmetric string theories in D < 10,''
Phys.\ Rev.\ Lett.\  {\bf 75}, 2264 (1995)
[arXiv:hep-th/9505054].

\bibitem{9506048}
S.~Chaudhuri and J.~Polchinski,
``Moduli space of CHL strings,''
Phys.\ Rev.\ D {\bf 52}, 7168 (1995)
[arXiv:hep-th/9506048].

\bibitem{9508144}
S.~Chaudhuri and D.~A.~Lowe,
``Type IIA heterotic duals with maximal supersymmetry,''
Nucl.\ Phys.\ B {\bf 459}, 113 (1996)
[arXiv:hep-th/9508144].

\bibitem{9508154}
P.~S.~Aspinwall,
``Some relationships between dualities in string theory,''
Nucl.\ Phys.\ Proc.\ Suppl.\  {\bf 46}, 30 (1996)
[arXiv:hep-th/9508154].

\bibitem{9903163}
J.~Maldacena,  G.~Moore and A.~Strominger,
``Counting BPS blackholes in toroidal type II string theory,''
arXiv:hep-th/9903163.

\bibitem{0005003}
  R.~Dijkgraaf, J.~M.~Maldacena, G.~W.~Moore, E.~P.~Verlinde,
  ``A Black hole Farey tail,''
  [hep-th/0005003].

\bibitem{0005017}
  S.~Carlip,
  ``Logarithmic corrections to black hole entropy from the Cardy formula,''
  Class.\ Quant.\ Grav.\  {\bf 17}, 4175 (2000)
  [arXiv:gr-qc/0005017].

\bibitem{9512078}
  C.~Vafa,
  ``Instantons on D-branes,''
  Nucl.\ Phys.\  {\bf B463}, 435-442 (1996).
  [hep-th/9512078].

\bibitem{9604042}
  J.~M.~Maldacena, L.~Susskind,
  ``D-branes and fat black holes,''
  Nucl.\ Phys.\  {\bf B475}, 679-690 (1996).
  [hep-th/9604042].

\bibitem{9601152}
  S.~R.~Das, S.~D.~Mathur,
  ``Excitations of D strings, entropy and duality,''
  Phys.\ Lett.\  {\bf B375}, 103-110 (1996).
  [hep-th/9601152].

\bibitem{9407001}
  S.~N.~Solodukhin,
  ``The Conical singularity and quantum corrections to entropy of black hole,''
  Phys.\ Rev.\  D {\bf 51}, 609 (1995)
  [arXiv:hep-th/9407001].

\bibitem{9408068}
  S.~N.~Solodukhin,
  ``On 'Nongeometric' contribution 
  to the entropy of black hole due to quantum
  corrections,''
  Phys.\ Rev.\  D {\bf 51}, 618 (1995)
  [arXiv:hep-th/9408068].


\bibitem{9412161}
  D.~V.~Fursaev,
  ``Temperature And Entropy Of A 
  Quantum Black Hole And Conformal Anomaly,''
  Phys.\ Rev.\  D {\bf 51}, 5352 (1995)
  [arXiv:hep-th/9412161].
  
\bibitem{9604118}
  R.~B.~Mann and S.~N.~Solodukhin,
  ``Conical geometry and quantum entropy of a charged Kerr black hole,''
  Phys.\ Rev.\  D {\bf 54}, 3932 (1996)
  [arXiv:hep-th/9604118].

\bibitem{9709064}
  R.~B.~Mann and S.~N.~Solodukhin,
  ``Universality of quantum entropy for extreme black holes,''
  Nucl.\ Phys.\  B {\bf 523}, 293 (1998)
  [arXiv:hep-th/9709064].

\bibitem{0002040}
  R.~K.~Kaul and P.~Majumdar,
  ``Logarithmic correction to the Bekenstein-Hawking entropy,''
  Phys.\ Rev.\ Lett.\  {\bf 84}, 5255 (2000)
  [arXiv:gr-qc/0002040].

\bibitem{0104010}
  T.~R.~Govindarajan, R.~K.~Kaul and V.~Suneeta,
  ``Logarithmic correction to the 
  Bekenstein-Hawking entropy of the BTZ  black
  hole,''
  Class.\ Quant.\ Grav.\  {\bf 18}, 2877 (2001)
  [arXiv:gr-qc/0104010].

\bibitem{0112041}
  K.~S.~Gupta, S.~Sen,
  ``Further evidence for the conformal structure of a Schwarzschild black hole in an algebraic approach,''
  Phys.\ Lett.\  {\bf B526}, 121-126 (2002).
  [hep-th/0112041].


\bibitem{0406044}
  A.~J.~M.~Medved,
  ``A comment on black hole entropy or why Nature abhors a logarithm,''
  Class.\ Quant.\ Grav.\  {\bf 22}, 133 (2005)
  [arXiv:gr-qc/0406044].

\bibitem{0409024}
  D.~N.~Page,
  ``Hawking radiation and black hole thermodynamics,''
  New J.\ Phys.\  {\bf 7}, 203 (2005)
  [arXiv:hep-th/0409024].


\bibitem{0805.2220}
  R.~Banerjee and B.~R.~Majhi,
  ``Quantum Tunneling Beyond Semiclassical Approximation,''
  JHEP {\bf 0806}, 095 (2008)
  [arXiv:0805.2220 [hep-th]].

\bibitem{0808.3688}
  R.~Banerjee and B.~R.~Majhi,
  ``Quantum Tunneling, Trace Anomaly and Effective Metric,''
  Phys.\ Lett.\  B {\bf 674}, 218 (2009)
  [arXiv:0808.3688 [hep-th]].

\bibitem{0809.1508}
  B.~R.~Majhi,
  ``Fermion Tunneling Beyond Semiclassical Approximation,''
  Phys.\ Rev.\  D {\bf 79}, 044005 (2009)
  [arXiv:0809.1508 [hep-th]].

\bibitem{0911.4379}
  R.~G.~Cai, L.~M.~Cao and N.~Ohta,
  ``Black Holes in Gravity with 
  Conformal Anomaly and Logarithmic Term in Black
  Hole Entropy,''
  arXiv:0911.4379 [hep-th].


\bibitem{1003.1083}
  R.~Aros, D.~E.~Diaz and A.~Montecinos,
  ``Logarithmic correction to BH entropy as Noether charge,''
  arXiv:1003.1083 [hep-th].

\bibitem{1008.4314}
  S.~N.~Solodukhin,
  ``Entanglement entropy of round spheres,''
  Phys.\ Lett.\  {\bf B693}, 605-608 (2010).
  [arXiv:1008.4314 [hep-th]].

\bibitem{1104.3712}
  S.~N.~Solodukhin,
  ``Entanglement entropy of black holes,''
  [arXiv:1104.3712 [hep-th]].

\bibitem{gibbhaw}
  G.~W.~Gibbons, S.~W.~Hawking,
  ``Action Integrals and Partition Functions in Quantum Gravity,''
  Phys.\ Rev.\  {\bf D15}, 2752-2756 (1977).

\bibitem{9905099}
  J.~M.~Bardeen and G.~T.~Horowitz,
  ``The extreme Kerr throat geometry: A vacuum analog of AdS(2) x S(2),''
  Phys.\ Rev.\ D {\bf 60}, 104030 (1999)
  [arXiv:hep-th/9905099].
  
\bibitem{0606244}
  D.~Astefanesei, K.~Goldstein, R.~P.~Jena, A.~Sen and S.~P.~Trivedi,
  ``Rotating attractors,''
  JHEP {\bf 0610}, 058 (2006)
  [arXiv:hep-th/0606244].

\bibitem{0705.4214}
  H.~K.~Kunduri, J.~Lucietti, H.~S.~Reall,
  ``Near-horizon symmetries of extremal black holes,''
  Class.\ Quant.\ Grav.\  {\bf 24}, 4169-4190 (2007).
  [arXiv:0705.4214 [hep-th]].


\bibitem{0806.2051}
  H.~K.~Kunduri, J.~Lucietti,
  ``A Classification of near-horizon geometries 
  of extremal vacuum black holes,''
  J.\ Math.\ Phys.\  {\bf 50}, 082502 (2009).
  [arXiv:0806.2051 [hep-th]].

\bibitem{0803.2998}
  P.~Figueras, H.~K.~Kunduri, J.~Lucietti, M.~Rangamani,
  ``Extremal vacuum black holes in higher dimensions,''
  Phys.\ Rev.\  {\bf D78}, 044042 (2008).
  [arXiv:0803.2998 [hep-th]].

\bibitem{0506177}
  A.~Sen,
  ``Black hole entropy function and the attractor mechanism in higher
  derivative gravity,''
  JHEP {\bf 0509}, 038 (2005)
  [arXiv:hep-th/0506177].

\bibitem{0905.2686}
  N.~Banerjee, S.~Banerjee, R.~Gupta, I.~Mandal and A.~Sen,
  ``Supersymmetry, Localization and Quantum Entropy Function,''
  arXiv:0905.2686 [hep-th].

\bibitem{1012.0265}
  A.~Dabholkar, J.~Gomes, S.~Murthy,
  ``Quantum black holes, localization and the topological string,''
  JHEP {\bf 1106}, 019 (2011).
  [arXiv:1012.0265 [hep-th]].


\bibitem{duffobs}
  M.~J.~Duff,
  ``Observations on Conformal Anomalies,''
  Nucl.\ Phys.\  {\bf B125}, 334 (1977).

\bibitem{christ-duff1}
  S.~M.~Christensen and M.~J.~Duff,
  ``New Gravitational Index Theorems And Supertheorems,''
  Nucl.\ Phys.\  B {\bf 154}, 301 (1979).

\bibitem{christ-duff2}
  S.~M.~Christensen and M.~J.~Duff,
  ``Quantizing Gravity With A Cosmological Constant,''
  Nucl.\ Phys.\  B {\bf 170}, 480 (1980).

\bibitem{duffnieu}
  M.~J.~Duff and P.~van Nieuwenhuizen,
  ``Quantum Inequivalence Of Different Field Representations,''
  Phys.\ Lett.\  B {\bf 94}, 179 (1980).


\bibitem{duffroc}
  S.~M.~Christensen, M.~J.~Duff, G.~W.~Gibbons and M.~Rocek,
  ``Vanishing One Loop Beta Function In Gauged N $>$ 4 Supergravity,''
  Phys.\ Rev.\ Lett.\  {\bf 45}, 161 (1980).

\bibitem{birrel}
N.~D.~Birrel and P.~C.~W.~Davis, Quantum Fields in Curved
Space, Cambridge University Press, New York, 1982.

\bibitem{gilkey}
P.~B.~Gilkey, 
``Invariance theory, the heat equation and the Atiyah-Singer index theorem,''
Publish or Perish Inc., USA (1984).

\bibitem{0306138}
  D.~V.~Vassilevich,
  ``Heat kernel expansion: User's manual,''
  Phys.\ Rept.\  {\bf 388}, 279 (2003)
  [arXiv:hep-th/0306138].

\bibitem{1009.4439}
  M.~J.~Duff, S.~Ferrara,
  ``Generalized mirror symmetry and trace anomalies,''
  Class.\ Quant.\ Grav.\  {\bf 28}, 065005 (2011).
  [arXiv:1009.4439 [hep-th]].

\bibitem{9912320}
  R.~C.~Henry,
  ``Kretschmann scalar for a kerr-newman black hole,''
  [astro-ph/9912320].

\bibitem{0302095}
  C.~Cherubini, D.~Bini, S.~Capozziello, R.~Ruffini,
  ``Second order scalar invariants of the Riemann tensor: Applications to black hole space-times,''
  Int.\ J.\ Mod.\ Phys.\  {\bf D11}, 827-841 (2002).
  [gr-qc/0302095].

\bibitem{0702072}
  A.~Castro, J.~L.~Davis, P.~Kraus, F.~Larsen,
  ``5D attractors with higher derivatives,''
  JHEP {\bf 0704}, 091 (2007).
  [hep-th/0702072].
  
\bibitem{0703087}
  A.~Castro, J.~L.~Davis, P.~Kraus, F.~Larsen,
  ``5D Black Holes and Strings with Higher Derivatives,''
  JHEP {\bf 0706}, 007 (2007).
  [hep-th/0703087].


\bibitem{0705.1847}
  A.~Castro, J.~L.~Davis, P.~Kraus, F.~Larsen,
  ``Precision Entropy of Spinning Black Holes,''
  JHEP {\bf 0709}, 003 (2007).
  [arXiv:0705.1847 [hep-th]].

\bibitem{0801.1863}
  A.~Castro, J.~L.~Davis, P.~Kraus, F.~Larsen,
  ``String Theory Effects on Five-Dimensional Black Hole Physics,''
  Int.\ J.\ Mod.\ Phys.\  {\bf A23}, 613-691 (2008).
  [arXiv:0801.1863 [hep-th]].

\bibitem{0807.0237}
  A.~Castro and S.~Murthy,
  ``Corrections to the statistical entropy of five dimensional black holes,''
  JHEP {\bf 0906}, 024 (2009)
  [arXiv:0807.0237 [hep-th]].

\bibitem{0910.4907}
  B.~de Wit, S.~Katmadas,
  ``Near-horizon analysis of D=5 BPS black holes and rings,''
  JHEP {\bf 1002}, 056 (2010).
  [arXiv:0910.4907 [hep-th]].

\bibitem{0912.0030}
  P.~D.~Prester,
  ``$AdS_3$ backgrounds from 10D effective
  action of heterotic string theory,''
  arXiv:0912.0030 [hep-th].

\bibitem{1001.1452}
  P.~D.~Prester,
  ``alpha'-corrections and heterotic black holes,''
  arXiv:1001.1452 [hep-th].

\bibitem{0710.3886}
  M.~Cvitan, P.~D.~Prester and A.~Ficnar,
  ``$alpha'^2$-corrections to extremal dyonic
black holes in heterotic string
  theory,''
  JHEP {\bf 0805}, 063 (2008)
  [arXiv:0710.3886 [hep-th]].

\bibitem{0809.4954}
  P.~D.~Prester and T.~Terzic,
  ``alpha'-exact entropies for BPS
and non-BPS extremal dyonic black holes in
  heterotic string theory from
ten-dimensional supersymmetry,''
  JHEP {\bf 0812}, 088 (2008)
  [arXiv:0809.4954 [hep-th]].

\bibitem{0611143}
  A.~Dabholkar, A.~Sen and S.~P.~Trivedi,
  ``Black hole microstates and attractor without supersymmetry,''
  JHEP {\bf 0701}, 096 (2007)
  [arXiv:hep-th/0611143].

\bibitem{0901.0359}
  N.~Banerjee, I.~Mandal, A.~Sen,
  ``Black Hole Hair Removal,''
  JHEP {\bf 0907}, 091 (2009).
  [arXiv:0901.0359 [hep-th]].


\bibitem{9204102}
  S.~Cecotti, P.~Fendley, K.~A.~Intriligator, C.~Vafa,
  ``A New supersymmetric index,''
  Nucl.\ Phys.\  {\bf B386}, 405-452 (1992).
  [hep-th/9204102].

\bibitem{9611205}
  C.~Bachas and E.~Kiritsis,
  ``F**4 terms in N = 4 string vacua,''
  Nucl.\ Phys.\ Proc.\ Suppl.\  {\bf 55B}, 194 (1997)
  [arXiv:hep-th/9611205].

\bibitem{9708062}
  A.~Gregori, E.~Kiritsis, C.~Kounnas, N.~A.~Obers,
  P.~M.~Petropoulos and B.~Pioline,
  ``R**2 corrections and non-perturbative
  dualities of N = 4 string ground
  states,''
  Nucl.\ Phys.\ B {\bf 510}, 423 (1998)
  [arXiv:hep-th/9708062].

\bibitem{0503217}
  D.~Gaiotto, A.~Strominger, X.~Yin,
  ``New connections between 4-D and 5-D black holes,''
  JHEP {\bf 0602}, 024 (2006).
  [hep-th/0503217].

\bibitem{0807.1314}
  N.~Banerjee,
  ``Subleading Correction to Statistical Entropy for BMPV Black Hole,''
  Phys.\ Rev.\  D {\bf 79}, 081501 (2009)
  [arXiv:0807.1314 [hep-th]].

\bibitem{0708.1270}
  A.~Sen,
  ``Black Hole Entropy Function,
Attractors and Precision Counting of
  Microstates,''
Gen.\ Rel.\ Grav.\  {\bf 40}, 2249 (2008)
  [arXiv:0708.1270 [hep-th]].
  

\bibitem{0907.0593}
  D.~P.~Jatkar, A.~Sen, Y.~K.~Srivastava,
  ``Black Hole Hair Removal: Non-linear Analysis,''
  JHEP {\bf 1002}, 038 (2010).
  [arXiv:0907.0593 [hep-th]].


\bibitem{camhig1}
  R.~Camporesi and A.~Higuchi,
  ``Spectral functions and zeta functions in hyperbolic spaces,''
  J.\ Math.\ Phys.\  {\bf 35}, 4217 (1994).

\bibitem{9505009}
  R.~Camporesi and A.~Higuchi,
  ``On The Eigen Functions Of 
  The Dirac Operator On Spheres And Real Hyperbolic
  Spaces,''
  J.\ Geom.\ Phys.\  {\bf 20}, 1 (1996)
  [arXiv:gr-qc/9505009].

\bibitem{0608021}
  C.~Beasley, D.~Gaiotto, M.~Guica, L.~Huang, 
  A.~Strominger and X.~Yin,
  ``Why Z(BH) = |Z(top)|**2,''
  arXiv:hep-th/0608021.


\end{thebibliography}
\end{document}